\documentclass[pdflatex,sn-basic]{sn-jnl}

\usepackage{amsmath}
\usepackage{amssymb}
\usepackage{booktabs}
\usepackage{multirow}
\usepackage{array}
\usepackage{xcolor}
\usepackage{listings}
\usepackage{algorithm}
\usepackage{algpseudocode}
\usepackage{tikz}
\usepackage{pgfplots}
\pgfplotsset{compat=1.18}
\usepackage{subcaption}    
\usepackage{url}
\usepackage{tabularx}
\usepackage{rotating}
\usepackage{makecell}
\usepackage{adjustbox}
\usepgfplotslibrary{groupplots}
\usepackage{arydshln}

\usetikzlibrary{arrows.meta,positioning,fit,backgrounds,calc,shapes.geometric,decorations.pathreplacing}

\definecolor{wazuhblue}{RGB}{0,82,147}
\definecolor{kafkadark}{RGB}{0,105,92}
\definecolor{elasticpurple}{RGB}{80,55,150}
\definecolor{kibanagreen}{RGB}{0,125,105}
\definecolor{agentgray}{RGB}{85,85,90}
\definecolor{aimodulebg}{RGB}{255,250,235}
\definecolor{aiborder}{RGB}{175,115,0}
\definecolor{consumerblue}{RGB}{30,100,170}

\tikzset{
  agent/.style={rectangle,rounded corners=3pt,fill=agentgray!12,draw=agentgray!60,thick,
    minimum width=1.85cm,minimum height=0.7cm,font=\small\sffamily,
    text=agentgray!85!black,align=center},
  wazuhbox/.style={rectangle,rounded corners=5pt,fill=wazuhblue!12,draw=wazuhblue,
    line width=1.3pt,minimum width=2.5cm,minimum height=1.4cm,
    font=\small\sffamily\bfseries,text=wazuhblue,align=center},
  kafkabox/.style={rectangle,rounded corners=4pt,fill=kafkadark!12,draw=kafkadark,
    line width=1.2pt,minimum width=2.1cm,minimum height=0.75cm,
    font=\small\sffamily\bfseries,text=kafkadark!85!black,align=center},
  aicomp/.style={rectangle,rounded corners=4pt,fill=aiborder!10,draw=aiborder,
    line width=1.2pt,minimum width=2.1cm,minimum height=0.75cm,
    font=\small\sffamily\bfseries,text=aiborder!85!black,align=center},
  catboostbox/.style={rectangle,rounded corners=4pt,fill=consumerblue!10,draw=consumerblue,
    line width=1.2pt,minimum width=2.5cm,minimum height=1.05cm,
    font=\small\sffamily\bfseries,text=consumerblue!85!black,align=center},
  elasticbox/.style={rectangle,rounded corners=5pt,fill=elasticpurple!12,draw=elasticpurple,
    line width=1.3pt,minimum width=2.4cm,minimum height=0.85cm,
    font=\small\sffamily\bfseries,text=elasticpurple,align=center},
  kibanabox/.style={rectangle,rounded corners=5pt,fill=kibanagreen!12,draw=kibanagreen,
    line width=1.3pt,minimum width=2.4cm,minimum height=0.8cm,
    font=\small\sffamily\bfseries,text=kibanagreen!80!black,align=center},
  idxbox/.style={rectangle,rounded corners=3pt,fill=elasticpurple!6,
    draw=elasticpurple!45,dashed,thick,minimum width=1.95cm,minimum height=0.6cm,
    font=\scriptsize\ttfamily,text=elasticpurple!75!black,align=center},
  arr/.style={-{Stealth[length=5pt,width=4pt]},thick,draw=#1},
  dbl/.style={{Stealth[length=4pt]}-{Stealth[length=4pt]},thick,draw=#1,dashed},
  lbl/.style={font=\scriptsize\sffamily,text=#1,align=center}
}

\definecolor{attackred}{RGB}{198,40,40}
\definecolor{normgreen}{RGB}{46,125,50}
\definecolor{catblue}{RGB}{30,100,170}
\definecolor{kibanagreen}{RGB}{0,125,105}
\definecolor{goldborder}{RGB}{175,115,0}
\definecolor{goldbg}{RGB}{255,250,235}

\tikzset{
  atknode/.style={rectangle,rounded corners=3pt,fill=attackred!10,draw=attackred,thick,
    minimum width=2.6cm,minimum height=0.75cm,font=\footnotesize\sffamily,
    text=attackred!85!black,align=center},
  victimnode/.style={rectangle,rounded corners=5pt,fill=wazuhblue!12,draw=wazuhblue,line width=1.3pt,
    minimum width=3.2cm,minimum height=1.1cm,font=\small\sffamily\bfseries,text=wazuhblue,align=center},
  selmnode/.style={rectangle,rounded corners=3pt,fill=normgreen!12,draw=normgreen,thick,
    minimum width=2.6cm,minimum height=0.85cm,font=\footnotesize\sffamily\bfseries,
    text=normgreen!80!black,align=center},
  wmnode/.style={rectangle,rounded corners=4pt,fill=wazuhblue!12,draw=wazuhblue,line width=1.2pt,
    minimum width=3.0cm,minimum height=0.85cm,font=\small\sffamily\bfseries,text=wazuhblue,align=center},
  esnode/.style={rectangle,rounded corners=4pt,fill=elasticpurple!12,draw=elasticpurple,line width=1.2pt,
    minimum width=2.6cm,minimum height=0.75cm,font=\small\sffamily\bfseries,text=elasticpurple,align=center},
  gtboxnode/.style={rectangle,rounded corners=3pt,fill=goldbg,draw=goldborder,dashed,thick,
    minimum width=3.0cm,minimum height=0.75cm,font=\scriptsize\sffamily,text=goldborder!80!black,align=center},
  inputboxn/.style={rectangle,rounded corners=4pt,fill=catblue!12,draw=catblue,line width=1.2pt,
    minimum width=5.0cm,minimum height=0.9cm,font=\small\sffamily\bfseries,text=catblue,align=center},
  stageone/.style={rectangle,rounded corners=4pt,fill=catblue!10,draw=catblue,line width=1.3pt,
    minimum width=6.5cm,minimum height=1.0cm,font=\small\sffamily\bfseries,text=catblue,align=center},
  stagetwo/.style={rectangle,rounded corners=4pt,fill=attackred!10,draw=attackred,line width=1.3pt,
    minimum width=6.5cm,minimum height=1.0cm,font=\small\sffamily\bfseries,text=attackred,align=center},
  decnode/.style={diamond,aspect=2.5,fill=goldborder!10,draw=goldborder,line width=1.2pt,
    font=\small\sffamily\bfseries,text=goldborder!80!black,align=center},
  normoutn/.style={rectangle,rounded corners=3pt,fill=normgreen!12,draw=normgreen,thick,
    minimum width=3.0cm,minimum height=0.8cm,font=\small\sffamily\bfseries,
    text=normgreen!80!black,align=center},
  atkclassr/.style={rectangle,rounded corners=3pt,fill=attackred!10,draw=attackred,thick,
    minimum width=2.2cm,minimum height=0.9cm,font=\scriptsize\sffamily\bfseries,
    text=attackred!85!black,align=center},
  storeboxn/.style={rectangle,rounded corners=4pt,fill=elasticpurple!10,draw=elasticpurple,
    line width=1.2pt,minimum width=5.5cm,minimum height=0.8cm,font=\small\sffamily\bfseries,
    text=elasticpurple,align=center},
  evtoldf/.style={rectangle,rounded corners=2pt,fill=agentgray!10,draw=agentgray!60,thick,
    minimum width=1.4cm,minimum height=0.65cm,font=\small\sffamily,align=center},
  evtcurf/.style={rectangle,rounded corners=3pt,fill=catblue!15,draw=catblue,line width=1.3pt,
    minimum width=1.6cm,minimum height=0.8cm,font=\small\sffamily\bfseries,text=catblue,align=center},
  featboxf/.style={rectangle,rounded corners=3pt,fill=#1!10,draw=#1,thick,
    minimum width=4.5cm,minimum height=1.2cm,font=\small\sffamily,text=#1!80!black,align=center},
  baseboxf/.style={rectangle,rounded corners=3pt,fill=catblue!10,draw=catblue,thick,
    minimum width=14.5cm,minimum height=1.1cm,font=\small\sffamily,text=catblue!80!black,align=center},
  vecboxf/.style={rectangle,rounded corners=4pt,fill=goldbg,draw=goldborder,line width=1.4pt,
    minimum width=7cm,minimum height=0.9cm,font=\small\sffamily\bfseries,
    text=goldborder!80!black,align=center},
  procn/.style={rectangle,rounded corners=4pt,fill=#1!10,draw=#1,line width=1.2pt,
    minimum width=3.8cm,minimum height=0.9cm,font=\small\sffamily\bfseries,
    text=#1!85!black,align=center},
  decnoder/.style={diamond,aspect=2.2,fill=goldbg,draw=goldborder,line width=1.2pt,
    font=\small\sffamily\bfseries,text=goldborder!80!black,align=center},
  mainboxd/.style={rectangle,rounded corners=4pt,fill=#1!10,draw=#1,line width=1.3pt,
    minimum width=4.8cm,minimum height=0.9cm,font=\small\sffamily\bfseries,
    text=#1!85!black,align=center},
  splitboxd/.style={rectangle,rounded corners=3pt,fill=#1!10,draw=#1,thick,
    minimum width=3.5cm,minimum height=1.0cm,font=\small\sffamily,text=#1!80!black,align=center},
  classboxd/.style={rectangle,rounded corners=2pt,fill=#1!10,draw=#1,thick,
    minimum width=2.1cm,minimum height=0.8cm,font=\scriptsize\sffamily,text=#1!80!black,align=center}
}
\begin{document}

\title[Context-Aware Web Attack Detection in Open-Source SIEM Systems]{Context-Aware Web Attack Detection in Open-Source SIEM Systems via MITRE ATT\&CK-Enriched Behavioral Profiling}

\author*[1]{\fnm{Badr} \sur{Alboushy}}\email{badr.alboushy@hiast.edu.sy}

\author[2]{\fnm{Assef} \sur{Jafar}}\email{assefjafar@gmail.com}

\author[3]{\fnm{Mohamad} \sur{Aljnidi}}\email{m-jnedi@aiu.edu.sy}

\author[1]{\fnm{Mohamad Bashar} \sur{Disoki}}\email{mohamadbashar.disoki@hiast.edu.sy}

\author[4]{\fnm{Aref} \sur{Shaheed}}\email{aref.shaheed@latakia-univ.edu.sy}

\affil*[1]{\orgname{Higher Institute for Applied Sciences and Technology (HIAST)}, \city{Damascus}, \country{Syria}}

\affil[2]{\orgname{Syrian Private University}, \city{Damascus}, \country{Syria}}

\affil[3]{\orgname{Arab International University}, \city{Damascus}, \country{Syria}}

\affil[4]{\orgname{Latakia University}, \city{Latakia}, \country{Syria}}

\abstract{Security Information and Event Management (SIEM) systems aggregate
log data from heterogeneous network devices and applications in order
to detect coordinated attacks that stateless per-event analysis
cannot reveal --- multi-step attack campaigns unfold across
sequences of events from a single source, requiring intra-source
temporal correlation rather than inter-source aggregation alone. Traditional rule-based correlation engines embedded in
open-source SIEM platforms are effective at detecting individually
recognisable anomalies yet struggle to classify multi-step, low-signal
web application attacks because they examine each event without
reference to the behavioural history of the originating host.

This paper presents \textsc{Smart-SIEM}, an AI module designed as a
modular enhancement for the open-source Wazuh SIEM platform.
The module requires configuring Wazuh rule levels to route events to the
classification pipeline; in regulated environments this configuration
step may require compliance review.
The module introduces two contributions. First, we define a per-source-IP
\emph{behavioural context vector} that summarises the most recent $N$
security events from the same host, encoding the distribution of HTTP
response-status buckets, the maximum rule activation count, and the
accumulation of MITRE ATT\&CK technique identifiers observed in prior
events. Second, we deploy a two-stage cascade in which Stage~1
distinguishes normal from malicious traffic and Stage~2 assigns one of
six fine-grained attack labels (SQL Injection, XSS, Web Vulnerability
Scanning, Brute Force, Broken Authentication, Sensitive Data Exposure).

We construct a purpose-built labelled dataset of 46,454 Wazuh security
events collected from a controlled testbed in which the OWASP Juice
Shop application serves as the victim, with simultaneous benign
Selenium-driven traffic and adversarial traffic generated by SQLMAP,
Acunetix, Burp Suite, and an XSS automation tool, each originating
from distinct IP addresses. Ground-truth labels are assigned
deterministically from IP identity and recorded attack timestamps.

A comparative evaluation of eight gradient boosting and baseline
algorithms on a single-session-per-class testbed (reported
F\textsubscript{1} values therefore represent an upper bound on
generalisation to unseen attack campaigns) demonstrates that
\emph{without} context features all algorithms converge to
$\approx$0.705 macro F\textsubscript{1}; \emph{with} context
features they rise to 0.947--0.967 (Stage~1) and 0.876--0.914
(Stage~2) --- an average improvement of $+0.254$ and $+0.324$
respectively. This algorithm-agnostic improvement confirms
that the behavioural context vector is the primary contribution.
A hybrid cascade combining LightGBM for Stage~1 and XGBoost for
Stage~2 achieves the best overall performance: F\textsubscript{1} of
0.967 (binary) and 0.914 (six-class). An ablation study over $N \in
\{3,\ldots,35\}$ identifies $N = 30$ as a practical operating point. Wazuh's
native rule engine detects 0\% of Brute Force and Broken
Authentication events; the AI module detects 100\% and 98.3\%
respectively. A self-adaptive retraining mechanism demonstrates
recovery from concept drift: F\textsubscript{1} drops from 0.905 to
0.465 when unseen attack types emerge, triggering retraining; partial recovery to 0.695 ($+0.099$) with Phase~2-only
retraining; the production-intended Phase~1+2 protocol recovers to
0.814 at the cost of a modest regression on originally-known classes.}


\maketitle

\section{Introduction}
\label{sec:intro}

The proliferation of web-facing services has made web application
attacks among the most prevalent and damaging threats facing
organisations today~\cite{gonzalez2021siem}. SQL injection, cross-site
scripting (XSS), credential stuffing, and automated vulnerability
scanning each manifest as extended campaigns rather than isolated
events: an attacker probes hundreds of endpoints, enumerates
directories, and escalates privilege across multiple sessions before
a successful breach is complete. Detecting such multi-step campaigns
requires correlating security events across time and across the
behavioural fingerprint of a given source address—a task that
exposes the fundamental limitation of purely rule-based SIEM
correlation engines.

Security Information and Event Management (SIEM) platforms occupy a
central role in enterprise security operations. They aggregate log data
from firewalls, intrusion detection systems (IDS), web servers, and
endpoints, normalise heterogeneous formats, apply correlation rules,
and raise alerts when rule conditions are satisfied. Open-source
platforms such as Wazuh~\cite{wazuh2022} provide a mature rule engine
aligned with standards including PCI~DSS, HIPAA, NIST~800-53, GDPR, and
the MITRE ATT\&CK framework~\cite{strom2018attack}. Yet the rule-engine
paradigm has three well-known weaknesses when confronted with
sophisticated web attacks: (1)~rules depend on expert-curated
signatures that cannot anticipate zero-day or polymorphic payloads;
(2)~each event is evaluated independently, so gradual reconnaissance
patterns that unfold over dozens of low-severity events are missed;
and (3)~the false positive burden is high because many benign
activities trigger the same low-level rule groups used as indicators
of attack.

Machine learning offers a complementary paradigm. Rather than matching
against fixed signatures, a trained classifier can learn statistical
boundaries between normal and malicious event sequences. However,
prior ML-based intrusion detection research largely targets
network-packet datasets (KDD\,99, NSL-KDD, CICIDS) rather than SIEM
log events, and almost universally treats each record as an
independent observation rather than as part of an ongoing session or
campaign~\cite{tavallaee2009nsl,sharafaldin2018cicids}.

This paper bridges that gap with three concrete contributions:

\begin{enumerate}

  \item \textbf{A contextual behavioural feature set for SIEM events.}
  For each incoming security event we construct a feature vector that
  aggregates the preceding $N$ events from the same source IP,
  encoding HTTP response-code distributions, peak rule activation
  frequency, and the cumulative count of each MITRE ATT\&CK technique
  identifier observed in the history window. This transforms a
  stateless event classifier into a session-aware detector.

  \item \textbf{A two-stage hybrid cascade classifier.}
  Stage~1 (LightGBM) makes a binary NORMAL/ATTACK decision. Only events
  flagged as ATTACK proceed to Stage~2 (XGBoost), which resolves the
  fine-grained attack category. A comparative study of eight algorithms
  demonstrates that the context vector improves \emph{all} tested
  algorithms by $+0.25$--$+0.35$ F\textsubscript{1}, establishing
  the context features rather than any specific algorithm as the
  primary contribution.

  \item \textbf{A self-adaptive retraining mechanism.}
  The system maintains a labelled knowledge base in Elasticsearch.
  Security analysts can add or relabel events; the system automatically
  evaluates current model accuracy against the knowledge base and
  retrains when accuracy falls below 90\%, allowing the deployed
  classifier to evolve with the monitored environment.

\end{enumerate}

The remainder of this paper is organised as follows.
Section~\ref{sec:related} reviews related work.
Section~\ref{sec:background} provides background on SIEM architecture,
gradient boosting classifiers, and SMOTE-NC.
Section~\ref{sec:system} describes the \textsc{Smart-SIEM} architecture.
Section~\ref{sec:dataset} details the dataset construction methodology.
Section~\ref{sec:features} defines the contextual feature engineering
procedure.
Section~\ref{sec:model} presents the cascaded classifier design and
training protocol.
Section~\ref{sec:results} reports experimental results and ablation
studies.
Section~\ref{sec:discussion} discusses implications and limitations.
Section~\ref{sec:conclusion} concludes.

\section{Related Work}
\label{sec:related}

\subsection{Rule-Based SIEM Correlation}

Security Information and Event Management (SIEM) systems arose from
the convergence of Security Information Management (SIM) and Security
Event Management (SEM), and their operational role in enterprise
security operations has been well established for over a
decade~\cite{miller2010siem,bhatt2014siem}.
The foundational purpose of a SIEM is to correlate events
from disparate sources in order to infer higher-level security states
that no single event reveals in isolation~\cite{muller2009correlation}.
\citet{chuvakin2012log} provide a comprehensive treatment
of log management, emphasising that the quality and completeness of
collected log data is a prerequisite for any correlation-based detection
strategy.
Rule-based correlation engines encode this knowledge as
condition-action pairs: when a set of logical predicates over event
fields is satisfied within a configurable time window, an alert is
raised~\cite{miller2010siem}.
Agrawal and Makwana~\cite{agrawal2015siem} survey the critical
capabilities of SIEM platforms and conclude that rule quality and
maintenance burden are the primary bottlenecks to accurate detection
in practice.
\citet{gonzalez2021siem} compare eight
commercial and open-source SIEM products (ArcSight, QRadar, McAfee
SIEM, LogRhythm, USM-OSSIM, RSA NetWitness, Splunk, SolarWinds) across
seventeen capability dimensions and find that open-source alternatives
lag commercial offerings primarily in data analytics, User and Entity
Behaviour Analytics (UEBA), and risk-analysis depth---capabilities that
machine learning can address without commercial licensing costs.
Despite their widespread adoption, rule-based engines face three
structural limitations: they depend on expert-curated signatures that
cannot anticipate zero-day or polymorphic
payloads~\cite{stallings2017security}; each event is evaluated
independently, so gradual reconnaissance campaigns that unfold over
dozens of low-severity events remain invisible; and the false positive
burden is high because broad rule groups match many benign
activities~\cite{garcia2009anomaly}.

\subsection{ML-Based Intrusion Detection}

The application of machine learning to intrusion detection has a long
history, tracing back to the seminal statistical anomaly model of
\citet{denning1987ids}, which established the conceptual
foundation for behaviour-based detection.
\citet{liao2013ids} and \citet{buczak2016ids}
provide extensive surveys of the subsequent literature, cataloguing
techniques ranging from na\"ive Bayes and support vector machines to
neural networks and ensemble methods.
The benchmark datasets that underpin much of this literature---including
KDD\,99, its refined successor NSL-KDD~\cite{tavallaee2009nsl}, and the
more recent CICIDS-2017~\cite{sharafaldin2018cicids}---represent
network-flow features rather than SIEM-normalised security events,
which limits their direct applicability inside a SIEM pipeline.
\citet{sommer2010outside} articulate a fundamental
tension: the high-dimensional, imbalanced, and concept-drifting nature
of real network traffic makes the standard evaluation assumptions of
machine learning research---closed-world classifiers, stationary
distributions, balanced classes---frequently invalid in operational
security settings.
Ensemble methods, in particular gradient boosting
variants~\cite{chen2016xgboost,ke2017lightgbm,breiman2001rf}, consistently
dominate classification benchmarks owing to their robustness to
irrelevant features and their ability to model non-linear interaction
effects without manual feature engineering.
Deep learning approaches---LSTM networks, convolutional models, and
attention-based architectures---have also shown strong performance on
sequential network data~\cite{ferrag2020deep,vinayakumar2019deep},
although they typically require larger labelled datasets and offer
less interpretability than gradient boosting on tabular inputs.
\citet{chandola2009anomaly} provide a unifying
taxonomy of anomaly detection methods and highlight that the scarcity
of labelled anomalies, combined with severe class imbalance, remains
a persistent obstacle to supervised detection---a challenge directly
addressed in this work through SMOTE-NC
oversampling~\cite{chawla2002smote,fernandez2018smote,he2009imbalanced,haixiang2017learning}.

\subsection{Web Application Attack Detection}

Web application attacks represent a distinct sub-domain of intrusion
detection because the relevant signals appear at the HTTP application
layer rather than in network packet headers.
SQL injection---the systematic injection of database commands through
user-supplied input fields---and cross-site scripting (XSS)---the
injection of executable client-side
code---remain among the most prevalent and damaging attack
vectors~\cite{halfond2006sql,grossman2007xss,owasp_top10}.
The OWASP Web Security Testing Guide~\cite{owasp_testing} catalogs
a broader taxonomy of web vulnerabilities and provides a reference
framework widely adopted in both academic and industrial testing.
Detecting these attacks from server-side logs is challenging because
individual malicious requests are often syntactically indistinguishable
from benign ones; meaningful signal emerges only when considering
the \emph{sequence} of requests from a given source
address~\cite{nisioti2018intrusion,chandola2009anomaly}.
The MITRE ATT\&CK framework~\cite{strom2018attack} provides a
structured vocabulary of adversary techniques, and recent work on
cyber threat intelligence extraction~\cite{husari2017ttpdrill} and
threat hunting~\cite{milajerdi2019poirot} has demonstrated that
ATT\&CK technique identifiers can serve as high-level behavioural
signatures that discriminate attack campaigns across diverse data
sources.
To the best of our knowledge, no prior work has accumulated ATT\&CK
technique frequencies across a per-IP event history window within a
SIEM pipeline and used those cumulative counts as discriminative
classifier features.

\subsection{ML-Based SIEM Enhancement}

Several works have attempted to augment SIEM platforms with machine
learning.
\citet{veeramachaneni2016aiid} propose an
AI$^2$ framework combining unsupervised anomaly detection with analyst
feedback loops; their system operates on generic log streams rather
than SIEM-normalised event formats and requires sustained analyst
labelling effort to function effectively.
\citet{sarker2020intrudtree} apply decision tree ensembles
to log-derived features but treat each event as an independent
observation, discarding the temporal context that distinguishes
reconnaissance campaigns from isolated probes.
\citet{ring2019ids} survey network-based intrusion detection
datasets and highlight the near-complete absence of SIEM-native
labelled collections, while \citet{kwon2019anomaly} review
deep learning for log-based anomaly detection but focus on system logs
rather than security-event correlation.
\citet{creech2014semantic} demonstrate that semantic
features derived from system call sequences substantially improve
detection of host-level intrusions, providing an analogy for the
hypothesis explored here: that behavioural \emph{sequences} of SIEM
events carry information that point-in-time features do not.
Interpretability of learned models is increasingly recognised as a
deployment requirement in security operations~\cite{bhatt2014siem}.
We address this through gain-based feature importance analysis
(Section~\ref{sec:results}), which confirms that context features
dominate Stage~1 discrimination.
Concept drift~\cite{gama2014concept,lu2019concept,yang2021cade,pendlebury2019tesseract}
motivates the self-adaptive retraining mechanism described in
Section~\ref{sec:system}.
Table~\ref{tab:related_work} situates \textsc{Smart-SIEM} against the
most closely related prior systems on six dimensions relevant to
operational SIEM deployment.

\begin{table*}[htbp]
\centering
\caption{Comparison of representative ML-based intrusion detection and
SIEM-augmentation systems. \checkmark~=~explicitly present;
$\circ$~=~partial or indirect; $\times$~=~absent.
\textit{Context features}: whether the system aggregates
temporal behavioural signals from prior events.
\textit{ATT\&CK}: whether MITRE ATT\&CK technique identifiers are
used as input features.
\textit{Drift}: whether the system includes a mechanism to adapt to
concept drift without full retraining.
\textit{SIEM platform}: whether the system integrates natively with
a deployed SIEM product.}

\label{tab:related_work}
\begin{adjustbox}{width=\textwidth,center}
\scriptsize
\setlength{\tabcolsep}{5pt}
\renewcommand{\arraystretch}{1.2}
\begin{tabular}{@{} l c l l c l c l @{}}
\toprule
\textbf{System} & \textbf{Year} & \textbf{Data Format}
  & \textbf{Context Features}
  & \textbf{ATT\&CK} & \textbf{Drift}
  & \textbf{SIEM} & \textbf{Task} \\
\midrule
AI$^2$~\cite{veeramachaneni2016aiid}
  & 2016 & Generic logs
  & $\circ$ time-window agg.
  & $\times$ & $\circ$ analyst feedback
  & $\times$ & Binary anomaly \\[2pt]
IntruDTree~\cite{sarker2020intrudtree}
  & 2020 & Network flows
  & $\times$ per-event only
  & $\times$ & $\times$ none
  & $\times$ & Multi-class IDS \\[2pt]
DeepLog~\cite{ferrag2020deep}
  & 2020 & System logs
  & $\circ$ LSTM (implicit)
  & $\times$ & $\circ$ incremental
  & $\times$ & Binary anomaly \\[2pt]
\citet{nisioti2018intrusion}
  & 2018 & IDS alert streams
  & $\circ$ multi-stage corr.
  & $\times$ & $\times$ none
  & $\circ$ generic & Attack attribution \\[2pt]
CADE~\cite{yang2021cade}
  & 2021 & Network flows
  & $\times$ per-event only
  & $\times$ & $\checkmark$ autoencoder
  & $\times$ & Anomaly~+~drift \\
\midrule
\textbf{Smart-SIEM (ours)}
  & \textbf{2022} & \textbf{Wazuh SIEM events}
  & $\checkmark$ \textbf{per-IP $N$=30 hist.}
  & $\checkmark$ \textbf{7 tech.}
  & $\checkmark$ \textbf{threshold}
  & $\checkmark$ \textbf{Wazuh}
  & \textbf{2-stage cascade} \\
\bottomrule
\end{tabular}
\end{adjustbox}
\renewcommand{\arraystretch}{1}
\end{table*}

Three differentiators are evident. First, no prior system operates
on Wazuh-normalised security events; all existing work uses network
flow datasets (NSL-KDD, CICIDS) or generic log streams that do not
preserve compliance mappings, MITRE ATT\&CK identifiers, or the
rule-correlation metadata native to SIEM platforms.
Second, the context vector in \textsc{Smart-SIEM} is the only
proposal to use \emph{cumulative MITRE ATT\&CK technique counts}
as explicit contextual features --- existing context-aware approaches
(AI$^2$, DeepLog) aggregate raw signal statistics without the
threat-framework semantics.
Third, the self-adaptive retraining loop in \textsc{Smart-SIEM}
is the only drift-adaptation mechanism that operates at the
knowledge-base accuracy level, allowing non-expert operators to
trigger retraining by labelling a small set of analyst-reviewed
events, without requiring a separate drift-detection
model~\cite{yang2021cade,pendlebury2019tesseract}.

\section{Background}
\label{sec:background}

\subsection{SIEM Architecture}

A SIEM platform comprises six functional components~\cite{miller2010siem,bhatt2014siem,chuvakin2012log}:
(1) \emph{data sources}—network devices, servers, and applications
that generate log records; (2) \emph{log collection}—agents or syslog
receivers that transport events to a centralised processor; (3)
\emph{parsing and normalisation}—extraction of structured fields from
free-form log text and mapping to a canonical schema; (4) \emph{correlation
engine}—rule-based or analytics-driven logic that identifies patterns
across multiple events; (5) \emph{log storage}—indexed persistence for
forensic analysis and compliance reporting; and (6) \emph{monitoring
and visualisation}—dashboards and alerting interfaces for security
analysts.

In Wazuh, the correlation engine applies XML-encoded rules to decoded
event fields. Each rule carries a severity \texttt{level} (0--16), a
\texttt{frequency} counter that tracks how many times it has fired
within a rolling time window, compliance mappings (PCI~DSS, HIPAA,
NIST, GDPR), and a MITRE ATT\&CK technique identifier. Events that
match a rule with \texttt{level} $> 0$ are forwarded to Elasticsearch
via Filebeat, where they are stored as JSON documents in the
\texttt{wazuh-alerts-*} index pattern.

\subsection{Gradient Boosting Classifiers}

The proposed hybrid uses LightGBM~\cite{ke2017lightgbm} for
Stage~1 and XGBoost~\cite{chen2016xgboost} for Stage~2;
CatBoost~\cite{prokhorenkova2018catboost,dorogush2018catboost}
is included as a comparison baseline. LightGBM uses a
histogram-based leaf-wise growth strategy that achieves fast
training on large datasets; it is the Stage~1 choice because
it achieves the highest binary F\textsubscript{1} (0.967) in our
comparison. XGBoost uses a level-wise growth strategy with a
regularised objective; it is the Stage~2 choice because it
achieves the highest multi-class F\textsubscript{1} (0.914).

CatBoost is a gradient-boosting
algorithm developed by Yandex that introduces two key innovations over
earlier approaches such as XGBoost~\cite{chen2016xgboost} and
LightGBM. First, it builds \emph{symmetric} (oblivious) decision trees
where every node at the same depth applies the same feature split,
which reduces prediction latency by transforming tree traversal into a
single vectorised comparison. Second, it employs
\emph{ordered target encoding} to eliminate prediction shift—a
systematic bias that arises in standard target encoding when the
statistic for a category is computed from examples that include the
current one. LightGBM~\cite{ke2017lightgbm} uses gradient-based one-side sampling (GOSS) and
exclusive feature bundling (EFB) to reduce training time while maintaining accuracy.
XGBoost~\cite{chen2016xgboost} uses second-order gradient statistics and
regularised tree learning with column sub-sampling.
Among the evaluated algorithms, CatBoost accepts categorical features in their raw string
form without the need for one-hot encoding, which makes it
well-suited for SIEM data where many fields (rule groups, compliance
tags, MITRE identifiers) are naturally categorical with high
cardinality.

\subsection{SMOTE-NC}

Class imbalance is endemic in security datasets~\cite{haixiang2017learning,he2009imbalanced} and SIEM deployments are no exception: normal traffic vastly
outnumbers labelled attack events in real deployments. The Synthetic
Minority Over-sampling Technique (SMOTE)~\cite{chawla2002smote}
addresses this by generating synthetic minority-class instances by
interpolating between each minority sample and one of its
$k$-nearest neighbours in feature space. SMOTE-NC (Nominal and
Continuous) extends SMOTE to mixed categorical-numerical
datasets~\cite{chawla2002smote}: for categorical features, the new
instance inherits the most frequent value among the $k$ neighbours,
while the standard Euclidean distance is augmented by the median
standard deviation of all numerical columns to account for categorical
mismatches. This prevents the nonsensical interpolation that naive
numeric encoding of categories would produce.

\section{System Architecture}
\label{sec:system}

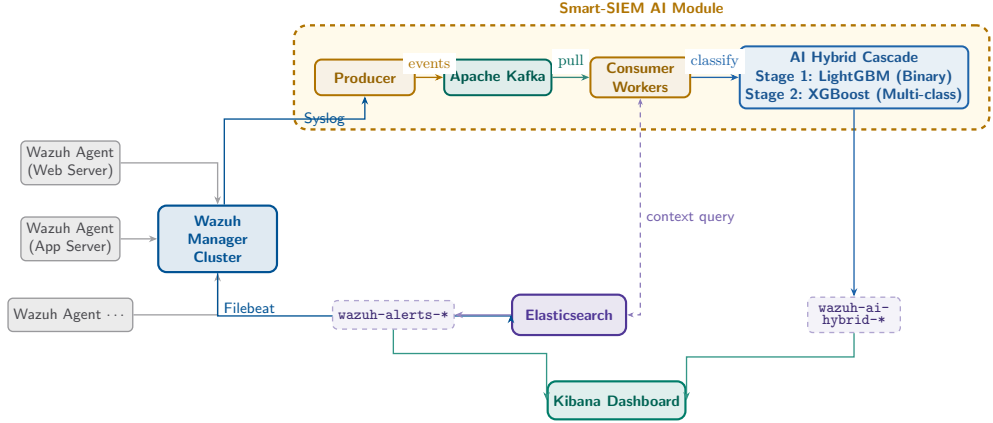
\begin{figure}[htbp]
\centering
\resizebox{\textwidth}{!}{%
\begin{tikzpicture}
\node[agent] (a1) at (0, 3.2) {Wazuh Agent\\(Web Server)};
\node[agent] (a2) at (0, 1.6) {Wazuh Agent\\(App Server)};
\node[agent] (a3) at (0, 0.0) {Wazuh Agent $\cdots$};
\node[wazuhbox] (wm) at (3.1, 1.6) {Wazuh\\Manager\\Cluster};
\draw[arr=agentgray!55] (a1.east) -| (wm.north);
\draw[arr=agentgray!55] (a2.east) -- (wm.west);
\draw[arr=agentgray!55] (a3.east) -| (wm.south);
\node[aicomp]      (prod)  at (6.2,  5.0) {Producer};
\node[kafkabox]    (kafka) at (9.0,  5.0) {Apache Kafka};
\node[aicomp]      (cons)  at (12.0, 5.0) {Consumer\\Workers};
\node[catboostbox] (cb)    at (16.5, 5.0) {AI Hybrid Cascade\\Stage~1: LightGBM (Binary)\\Stage~2: XGBoost (Multi-class)};
\draw[arr=aiborder] (prod.east) -- (kafka.west)
    node[font=\scriptsize, text=aiborder, fill=white,
         midway, above, yshift=2pt]{events};
\draw[arr=kafkadark!80] (kafka.east) -- (cons.west)
    node[font=\scriptsize, text=kafkadark, fill=white,
         midway, above, yshift=2pt]{pull};
\draw[arr=consumerblue] (cons.east) -- (cb.west)
    node[font=\scriptsize, text=consumerblue, fill=white,
         midway, above, yshift=2pt]{classify};
\begin{scope}[on background layer]
  \node[
    fill=aimodulebg, draw=aiborder, dashed, line width=1.4pt,
    rounded corners=8pt, inner sep=12pt,
    fit=(prod)(kafka)(cons)(cb)
  ] (aimodule) {};
\end{scope}
\node[font=\small\sffamily\bfseries, text=aiborder]
  at (aimodule.north) [above=3pt] {Smart-SIEM AI Module};
\draw[arr=wazuhblue]
  ([xshift=4pt]wm.north) -- ++(0, 1.8) -|
  node[lbl=wazuhblue, near start, right, xshift=2pt]{\scriptsize Syslog}
  (prod.south);
\node[elasticbox] (es) at (10.5, 0.0) {Elasticsearch};
\draw[arr=wazuhblue]
  (wm.south) -- ++(0,-0.9)
  node[lbl=wazuhblue, right, yshift=5pt]{\scriptsize Filebeat}
  -| (es.west);
\draw[dbl=elasticpurple!65]
  (cons.south) -- ++(0,-1.2) -|
  node[lbl=elasticpurple!80, near end, right, yshift=4pt]{\scriptsize context query}
  (es.north -| {(12.0,0)}) |- (es.east);
\node[idxbox] (idx2) at (16.5, 0.0) {\texttt{wazuh-ai-}\\[-2pt]\texttt{hybrid-*}};
\draw[arr=consumerblue] (cb.south) -- (idx2.north);
\node[idxbox] (idx1) at (6.8, 0.0) {\texttt{wazuh-alerts-*}};
\draw[arr=elasticpurple!60] (es.west) -- (idx1.east);
\node[kibanabox] (kibana) at (11.5, -1.8) {Kibana Dashboard};
\draw[arr=kibanagreen!80] (idx1.south) -- ++(0,-0.5) -| (kibana.west);
\draw[arr=kibanagreen!80] (idx2.south) -- ++(0,-0.5) -| (kibana.east);
\end{tikzpicture}
} 
\caption{Architecture of \textsc{Smart-SIEM}. Security events produced
by Wazuh agents are processed by the Wazuh Manager cluster, which
forwards events simultaneously to Elasticsearch (standard
\texttt{wazuh-alerts-*} index via Filebeat) and to the AI module via
syslog. The AI module enqueues events in Apache Kafka; consumer workers
retrieve per-IP context from Elasticsearch and classify each event with
the hybrid cascade model. Classification results are written to a
dedicated \texttt{wazuh-ai-hybrid-*} index and visualised in a
custom Kibana dashboard.}
\label{fig:arch}
\end{figure}

\textsc{Smart-SIEM} is architected as a loosely coupled module that
attaches to an existing Wazuh deployment without modifying its core
components. Figure~\ref{fig:arch} shows the overall data flow.

\subsection{Wazuh Manager Cluster}

The Wazuh Manager cluster is configured to emit every security event
whose rule \texttt{level} exceeds zero as a syslog message directed to
the AI module's producer listener. To ensure that low-severity
(\texttt{level}~=~0) events—which Wazuh normally silences—also reach
the AI module, we applied a bulk rule transformation that promotes all
\texttt{level}~=~0 rules to \texttt{level}~=~1. These benign events
carry valuable negative-class signal and their rule-field metadata
(groups, MITRE identifiers, compliance tags) helps the classifier
characterise normal behaviour.

\subsection{Apache Kafka Message Queue}

Each Wazuh Manager node runs a Python \emph{producer} process that
parses the incoming syslog stream, deserialises the JSON payload, and
publishes the event to a Kafka topic. Kafka provides back-pressure
handling, at-least-once delivery guarantees, and horizontal fan-out to
multiple consumer groups, satisfying the scalability and
fault-tolerance requirements of a production SIEM pipeline.

\subsection{Consumer and Classification Pipeline}

Consumer workers pull messages from Kafka and execute the following
pipeline for each event:

\begin{enumerate}
  \item \textbf{Preprocessing.} Missing categorical fields are filled
  with empty strings; missing numerical fields receive default values
  (rule level: 3, firedtimes: 1). The \texttt{rule.mail} boolean is
  mapped to $\{0,1\}$.

  \item \textbf{Context retrieval.} The consumer queries Elasticsearch
  for the $N$ most recent events whose \texttt{data.srcip} matches the
  current event's source IP and whose timestamp precedes the current
  event. The context window $N$ is a configurable parameter; our
  ablation study (Section~\ref{sec:ablation}) identifies $N = 30$ as
  the optimal value.

  \item \textbf{Context feature construction.} From the retrieved
  history the consumer computes the context feature vector (defined
  formally in Section~\ref{sec:features}) and appends it to the
  current event's feature vector.

  \item \textbf{Stage~1 classification.} The enriched feature vector
  is passed to the LightGBM binary classifier (Stage~1). If the
  prediction is \texttt{NORMAL}, the event is written directly to the
  output index.

  \item \textbf{Stage~2 classification.} Events predicted as
  \texttt{ATTACK} by Stage~1 are passed to the multi-class XGBoost
  classifier (Model~2), which assigns one of six attack labels.

  \item \textbf{Result persistence.} The classified event, including
  both Stage~1 and Stage~2 labels, is written to the
  \texttt{wazuh-ai-hybrid-*} Elasticsearch index.
\end{enumerate}

\subsection{Self-Adaptive Retraining}

To maintain accuracy as the monitored environment evolves, the system
maintains a labelled knowledge base index
(\texttt{wazuh-ai-knowledge\_base}) in Elasticsearch. Analysts may add
new labelled events or correct existing labels via a dedicated
interface. Each day, the consumer checks classifier accuracy against
the full knowledge base. If macro-averaged accuracy falls below 90\%,
new models are trained on the updated knowledge base and deployed,
replacing the previous models. The 90\% threshold is configurable.

\begin{figure}[htbp]
\centering
\resizebox{0.60\textwidth}{!}{%
\begin{tikzpicture}[node distance=0.8cm]
\node[procn=catblue]       (prod)     {Hybrid Cascade Models\\(LightGBM Stage~1 $+$ XGBoost Stage~2)};
\node[procn=catblue,below=0.7cm of prod]  (classify) {Classify Incoming\\Security Events};
\node[procn=kibanagreen,below=0.7cm of classify] (dash) {Kibana Dashboard\\(Analyst Review Interface)};
\node[procn=elasticpurple,below=0.7cm of dash]   (kb)
    {Knowledge Base\\(\texttt{wazuh-ai-knowledge\_base})\\Analyst adds / corrects labels};
\node[procn=goldborder,below=0.7cm of kb]    (eval)
    {Daily Accuracy Evaluation\\(macro-averaged, vs.\ full KB)};
\node[decnoder,below=0.7cm of eval]          (dec)    {Accuracy\\$\geq 90\%$?};
\node[procn=normgreen,right=2.5cm of dec]    (cont)   {Continue --- no\\model update};
\node[procn=attackred,below=0.8cm of dec]    (retrain){Trigger Retraining\\on Updated KB};
\node[procn=normgreen,below=0.7cm of retrain](newmod) {Deploy New Models\\(replace production)};
\draw[-{Stealth[length=5pt]},thick,draw=catblue]       (prod)    -- (classify);
\draw[-{Stealth[length=5pt]},thick,draw=catblue]       (classify)-- (dash);
\draw[-{Stealth[length=5pt]},thick,draw=kibanagreen]   (dash)    -- (kb);
\draw[-{Stealth[length=5pt]},thick,draw=elasticpurple] (kb)      -- (eval);
\draw[-{Stealth[length=5pt]},thick,draw=goldborder]    (eval)    -- (dec);
\draw[-{Stealth[length=5pt]},thick,draw=normgreen,line width=1.3pt] (dec.east) -- (cont.west)
    node[font=\small\sffamily,text=normgreen,midway,above]{YES};
\draw[-{Stealth[length=5pt]},thick,draw=attackred,line width=1.3pt] (dec.south) -- (retrain.north)
    node[font=\small\sffamily,text=attackred,midway,right]{NO};
\draw[-{Stealth[length=5pt]},thick,draw=normgreen] (retrain) -- (newmod);
\draw[-{Stealth[length=5pt]},thick,draw=catblue,line width=1.3pt]
    (newmod.west) -- ++(-2.0,0) |- (prod.west);
\node[font=\scriptsize\sffamily,text=goldborder!80!black,align=center]
    at ([xshift=3.5cm]eval.east) {Threshold: 90\%\\(configurable)};
\end{tikzpicture}
}
\caption{Self-adaptive retraining loop.
The hybrid cascade models classify events continuously; results are surfaced in a
Kibana dashboard where security analysts can add or correct labels, which are stored in
the \texttt{wazuh-ai-knowledge\_base} Elasticsearch index.
Each day the system evaluates macro-averaged classifier accuracy against the full
knowledge base; if accuracy falls below the 90\% threshold, both Stage~1 and Stage~2
models are retrained on the updated knowledge base and the new models replace the
current production versions.}
\label{fig:retrain}
\end{figure}
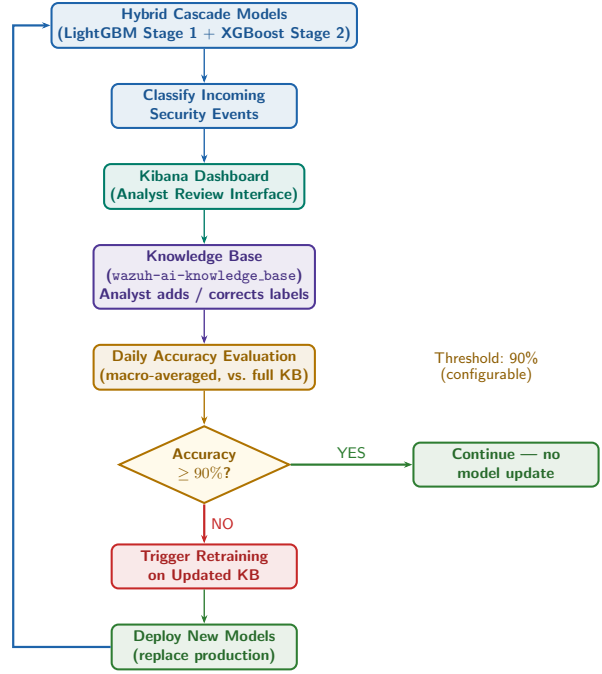

\section{Dataset Construction}
\label{sec:dataset}

\subsection{Testbed Configuration}

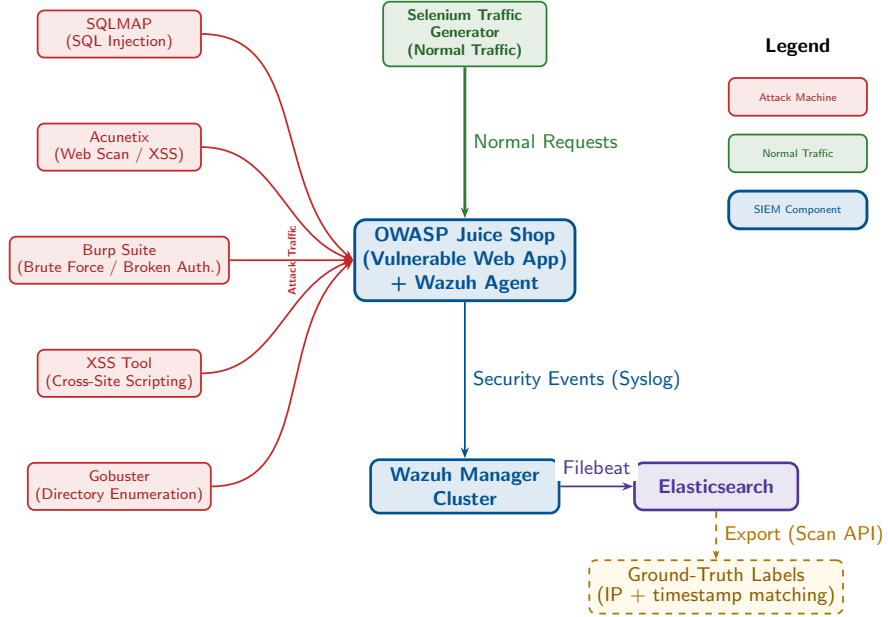
\begin{figure}[htbp]
\centering
\resizebox{0.90\textwidth}{!}{%
\begin{tikzpicture}
\node[atknode] (a1) at (0,  4.5) {SQLMAP\\(SQL Injection)};
\node[atknode] (a2) at (0,  2.7) {Acunetix\\(Web Scan / XSS)};
\node[atknode] (a3) at (0,  0.9) {Burp Suite\\(Brute Force / Broken Auth.)};
\node[atknode] (a4) at (0, -0.9) {XSS Tool\\(Cross-Site Scripting)};
\node[atknode] (a5) at (0, -2.7) {Gobuster\\(Directory Enumeration)};

\node[victimnode] (vic) at (5.5, 0.9)
    {OWASP Juice Shop\\(Vulnerable Web App)\\+ Wazuh Agent};

\node[selmnode] (sel) at (5.5, 4.5)
    {Selenium Traffic\\Generator\\(Normal Traffic)};

\draw[-{Stealth[length=5pt]},thick,draw=attackred] (a1.east) to[out=0,in=140] (vic.west);
\draw[-{Stealth[length=5pt]},thick,draw=attackred] (a2.east) to[out=0,in=160] (vic.west);
\draw[-{Stealth[length=5pt]},thick,draw=attackred] (a3.east) --               (vic.west);
\draw[-{Stealth[length=5pt]},thick,draw=attackred] (a4.east) to[out=0,in=200] (vic.west);
\draw[-{Stealth[length=5pt]},thick,draw=attackred] (a5.east) to[out=0,in=220] (vic.west);

\node[font=\tiny\sffamily\bfseries, text=attackred, rotate=90]
    at (2.75, 0.9) {Attack Traffic};

\draw[-{Stealth[length=5pt]},thick,draw=normgreen,line width=1.3pt]
    (sel.south) -- (vic.north)
    node[font=\scriptsize\sffamily, text=normgreen, midway, right]
    {Normal Requests};

\node[wmnode] (wm) at (5.5, -2.7) {Wazuh Manager\\Cluster};

\draw[-{Stealth[length=5pt]},thick,draw=wazuhblue]
    (vic.south) -- (wm.north)
    node[font=\scriptsize\sffamily, text=wazuhblue, midway, right]
    {Security Events (Syslog)};

\node[esnode] (es) at (9.5, -2.7) {Elasticsearch};

\draw[-{Stealth[length=5pt]},thick,draw=elasticpurple]
    (wm.east) -- (es.west)
    node[font=\scriptsize\sffamily, text=elasticpurple,
         fill=white, midway, above, yshift=2pt]{Filebeat};

\node[gtboxnode] (gt) at (9.5, -4.3)
    {Ground-Truth Labels\\(IP $+$ timestamp matching)};

\draw[-{Stealth[length=4pt]},thick,draw=goldborder,dashed]
    (es.south) -- (gt.north)
    node[font=\scriptsize\sffamily, text=goldborder, midway, right]
    {Export (Scan API)};

\node[font=\small\sffamily\bfseries] at (10.8, 4.3) {Legend};
\node[atknode,  minimum width=2.2cm, minimum height=0.6cm,
      font=\tiny\sffamily] at (10.8, 3.5) {Attack Machine};
\node[selmnode, minimum width=2.2cm, minimum height=0.6cm,
      font=\tiny\sffamily] at (10.8, 2.6) {Normal Traffic};
\node[victimnode, minimum width=2.2cm, minimum height=0.6cm,
      font=\tiny\sffamily] at (10.8, 1.7) {SIEM Component};

\end{tikzpicture}
}
\caption{Testbed configuration used for dataset collection.
Five physically separate attack machines run dedicated tools against the OWASP Juice Shop
victim server, while a Selenium script generates concurrent normal traffic from a separate IP.
The Wazuh Agent on the victim server forwards all security events to the Wazuh Manager
cluster, which persists them in Elasticsearch via Filebeat.
Ground-truth labels are assigned deterministically from IP identity and recorded
attack-session timestamps.}
\label{fig:testbed}
\end{figure}

Because no publicly available labelled SIEM dataset exists that
captures multi-class web application attacks in Wazuh's event format,
we constructed a purpose-built dataset using a controlled testbed
(Figure~\ref{fig:testbed}). The testbed consists of:

\begin{itemize}
  \item \textbf{Victim server.} OWASP Juice Shop~\cite{juiceshop}, a
  deliberately vulnerable Node.js e-commerce application that
  implements all categories of the OWASP Top~10. A Wazuh agent was
  deployed on this server and configured to monitor the application's
  Apache access log.

  \item \textbf{Normal traffic generator.} A Python script using the
  Selenium WebDriver library continuously submitted random legitimate
  requests (browse products, add to cart, register, login) to Juice
  Shop from a designated IP address throughout all experimental
  sessions.

  \item \textbf{Attack machines.} One or more physically separate
  machines with distinct IP addresses performed attack sessions using
  the tools listed in Table~\ref{tab:tools}.
\end{itemize}

\begin{tableorg}[htbp]
\centering
\caption{Attack tools used during dataset collection.}
\label{tab:tools}
\renewcommand{\arraystretch}{1.3}
\resizebox{\textwidth}{!}{%
\begin{tabular}{@{} l l l l @{}}
\toprule
\textbf{Tool} & \textbf{Attack Type} & \textbf{Mode} & \textbf{Ref.} \\
\midrule
SQLMAP              & SQL Injection                      & Auto.\ + manual  & \cite{sqlmap}    \\
Acunetix            & Web Vulnerability Scanning / XSS  & Automated        & \cite{acunetix}  \\
Burp Suite          & Brute Force / Broken Auth.        & Semi-automated   & \cite{burpsuite} \\
XSS automation tool & Cross-Site Scripting               & Automated        & ---              \\
Gobuster            & Web Scanning (dir.\ enum.)        & Automated        & \cite{gobuster}  \\
\bottomrule
\end{tabular}}
\renewcommand{\arraystretch}{1}
\end{tableorg}

\subsection{Ground-Truth Labelling}

For each attack session we recorded: (a) the source IP of the
attacking machine; (b) the precise start and end timestamps of the
attack. After data collection, Wazuh security events were exported
from Elasticsearch using the Scan API. Each event was assigned a
ground-truth label deterministically: events whose \texttt{data.srcip}
matches a known attack IP and whose \texttt{timestamp} falls within a
recorded attack window receive the corresponding attack class label;
all remaining events are labelled \texttt{NORMAL}. This procedure
avoids the ambiguity of manual event-by-event annotation and produces
a clean ground truth tied to observable network identity.

\subsection{Dataset Statistics}

\textbf{Two-sided balancing strategy.}
\texttt{SENSITIVE\_DATA\_EXPOSURE} dominates the raw dataset with
26,558 instances (57.2\%), while \texttt{BROKEN\_AUTHENTICATION}
has only 300 instances (0.6\%). We apply a two-sided balancing
strategy to the training split: minority classes are
\emph{oversampled} to 1,250 instances each via SMOTE-NC, while
the dominant class is \emph{randomly undersampled} to 1,250
instances. The 95\% reduction in \texttt{SENSITIVE\_DATA\_EXPOSURE}
training instances is intentional: all six attack classes are
operationally equally important to detect. The test set is
evaluated at its natural class proportions (no balancing),
providing an unbiased estimate of real-world performance.

The raw dataset contains 46,454 security events with 43 fields each.
Table~\ref{tab:class_distribution} shows the class distribution before
and after SMOTE-NC balancing of the training split.

\begin{tableorg}[htbp]
\centering
\caption{Class distribution of the raw dataset and the SMOTE-NC
balanced training set.}
\label{tab:class_distribution}
\renewcommand{\arraystretch}{1.25}
\resizebox{\textwidth}{!}{%
\begin{tabular}{@{} l rr rr @{}}
\toprule
\multirow{2}{*}{\textbf{Class}}
  & \multicolumn{2}{c}{\textbf{Raw Dataset}}
  & \multicolumn{2}{c}{\textbf{Training Set (SMOTE-NC)}} \\
\cmidrule(lr){2-3}\cmidrule(lr){4-5}
  & \textbf{Count} & \textbf{\%}
  & \textbf{Count} & \textbf{\%} \\
\midrule
\texttt{SENSITIVE\_DATA\_EXPOSURE} & 26{,}558 & 57.2 &  1{,}250 & 10.0 \\
\texttt{SQL\_INJECTION}            &  6{,}573 & 14.2 &  1{,}250 & 10.0 \\
\texttt{NORMAL}                    &  6{,}350 & 13.7 &  5{,}000 & 40.0 \\
\texttt{WEB\_SCAN}                 &  5{,}654 & 12.2 &  1{,}250 & 10.0 \\
\texttt{BRUTE\_FORCE}              &    702   &  1.5 &  1{,}250 & 10.0 \\
\texttt{XSS}                       &    317   &  0.7 &  1{,}250 & 10.0 \\
\texttt{BROKEN\_AUTHENTICATION}    &    300   &  0.6 &  1{,}250 & 10.0 \\
\midrule
\textbf{Total}
  & \textbf{46{,}454} & \textbf{100}
  & \textbf{12{,}500} & \textbf{100} \\
\bottomrule
\end{tabular}}
\renewcommand{\arraystretch}{1}
\end{tableorg}

The dominance of \texttt{SENSITIVE\_DATA\_EXPOSURE} in the raw data
is expected and reflects the operational behaviour of web scanning and
SQL injection tools: a single SQLMAP or Acunetix session generates
hundreds to thousands of HTTP requests, the majority of which trigger
Wazuh's sensitive-data-exposure rules (e.g., responses that include
stack traces, verbose error messages, or partial database query text).
This is a realistic distribution for the attack tools used; it is not
a data collection artefact.

\subsection{Train / Validation / Test Split}

Events are partitioned 64\%/16\%/20\% using stratified
random sampling to preserve class proportions across all six attack
classes (verified: the test set contains $19.9$--$20.0$\% of each
class). SMOTE-NC was applied
\emph{exclusively} to the training split; no synthetic samples appear
in the validation or test sets.

\textbf{Context feature construction and data split.}
Context features are computed on the complete event log sorted
chronologically by timestamp, \emph{before} any split.
For each event $e_i$, its context vector uses the $N=30$ most recent
prior events from the same source IP --- strictly enforced by the
dataset-construction code, which iterates backward through
earlier-timestamped rows only (no future events can enter the window).

Events are then partitioned into training (64\%), validation (16\%),
and test (20\%) using stratified random sampling, preserving class
proportions across all six attack classes (test set contains
$19.9$--$20.0$\% of each class, verified by inspection).

\begin{figure}[htbp]
\centering
\resizebox{\textwidth}{!}{%
\begin{tikzpicture}

\node[mainboxd=elasticpurple] (raw) at (0,0)
    {46,454 Wazuh Security Events\\(exported from Elasticsearch)};

\node[mainboxd=goldborder, below=0.7cm of raw] (lbl)
    {Deterministic Ground-Truth Labelling\\
     by Source IP $+$ Attack Timestamps};
\draw[-{Stealth[length=5pt]},thick,draw=elasticpurple] (raw) -- (lbl);

\node[classboxd=normgreen, below=1.0cm of lbl,xshift=-6.6cm] (c1)
    {NORMAL\\6,350 (13.7\%)};
\node[classboxd=attackred, below=1.0cm of lbl,xshift=-4.4cm] (c2)
    {Sensitive\\Data Exp.\\26,558};
\node[classboxd=attackred, below=1.0cm of lbl,xshift=-2.2cm] (c3)
    {SQL Inj.\\6,573};
\node[classboxd=attackred, below=1.0cm of lbl,xshift= 0.0cm] (c4)
    {Web Scan\\5,654};
\node[classboxd=attackred, below=1.0cm of lbl,xshift= 2.2cm] (c5)
    {Brute\\Force\\702};
\node[classboxd=attackred, below=1.0cm of lbl,xshift= 4.4cm] (c6)
    {XSS\\317};
\node[classboxd=attackred, below=1.0cm of lbl,xshift= 6.6cm] (c7)
    {Broken\\Auth.\\300};

\foreach \c in {c1,c2,c3,c4,c5,c6,c7}
    \draw[-{Stealth[length=4pt]},thin,draw=goldborder] (lbl.south) -- (\c.north);

\node[font=\scriptsize\sffamily\bfseries, text=goldborder,
      below=0.1cm of c4] (datasetlbl)
    {7-Class Labelled Dataset --- 46,454 events total};

\node[mainboxd=agentgray, below=0.9cm of datasetlbl] (split)
    {Stratified Random Event-Level Split};

\foreach \c in {c1,c2,c3,c4,c5,c6,c7}
    \draw[-{Stealth[length=4pt]},thin,draw=agentgray]
        (\c.south) -- (split.north);

\node[splitboxd=catblue,   below=1.0cm of split,xshift=-4.5cm] (train)
    {\textbf{Training Set}\\64\,\% --- 29,731 events};
\node[splitboxd=agentgray, below=1.0cm of split,xshift= 0.0cm] (val)
    {\textbf{Validation Set}\\16\,\% --- 7,432 events\\(no SMOTE applied)};
\node[splitboxd=agentgray, below=1.0cm of split,xshift= 4.5cm] (test)
    {\textbf{Test Set}\\20\,\% --- 9,232 events\\(no SMOTE applied)};

\draw[-{Stealth[length=5pt]},thick,draw=catblue]
    (split.south) -- ++(0,-0.5) -| (train.north);
\draw[-{Stealth[length=5pt]},thick,draw=agentgray]
    (split.south) -- ++(0,-0.5) -- (val.north);
\draw[-{Stealth[length=5pt]},thick,draw=agentgray]
    (split.south) -- ++(0,-0.5) -| (test.north);

\node[mainboxd=goldborder, below=0.8cm of train] (smote)
    {SMOTE-NC Balancing\\(training split only)};
\draw[-{Stealth[length=5pt]},thick,draw=catblue] (train) -- (smote);

\node[mainboxd=normgreen, below=0.8cm of smote] (balanced)
    {Balanced Training Set --- 12,500 events\\
     NORMAL: 5,000 \quad Each attack class: 1,250};
\draw[-{Stealth[length=5pt]},thick,draw=normgreen,line width=1.3pt]
    (smote) -- (balanced);

\draw[-{Stealth[length=4pt]},dashed,thin,draw=attackred]
    (val.south)  -- ++(0,-0.45);
\draw[-{Stealth[length=4pt]},dashed,thin,draw=attackred]
    (test.south) -- ++(0,-0.45);
\node[font=\scriptsize\sffamily\bfseries, text=attackred,
      below=0.5cm of val, xshift=2.25cm]
    {$\dagger$ evaluation only (no resampling) $\dagger$};

\end{tikzpicture}
}
\caption{Dataset construction and splitting pipeline.
Raw events are exported from Elasticsearch and assigned ground-truth labels
deterministically from source IP and attack-session timestamps.
The 46,454-event corpus is partitioned at session level into training (64\%),
validation (16\%), and test (20\%) splits.
SMOTE-NC oversampling is applied exclusively to the training split;
validation and test sets contain only real events at their natural class proportions.}
\label{fig:dataset}
\end{figure}
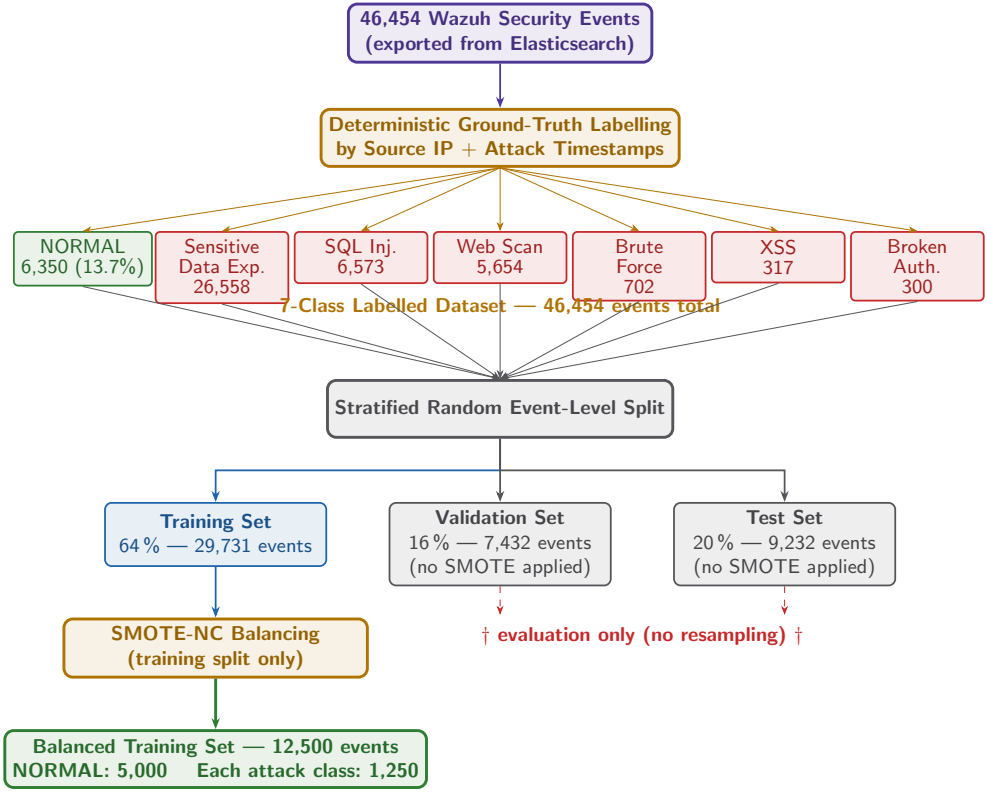

\textbf{Data isolation argument.}
Because context features are computed with strict temporal ordering,
a test-set event's context window may reference training-split events
(those that occurred earlier in time from the same source IP).
This does \emph{not} constitute methodological leakage for two reasons.
First, context features aggregate only \emph{behavioural signals}
--- HTTP status-code distributions and MITRE ATT\&CK technique
frequencies --- not class labels; no label from a training event
can contaminate a test event's feature vector.
Second, this construction accurately mirrors the production deployment,
where the Elasticsearch history query retrieves all chronologically
prior events from a source IP regardless of when the model was trained.
The reported test-set performance therefore represents a realistic
estimate of operational accuracy under genuine deployment conditions.
The seven source IPs correspond to the seven attack tools and
traffic sources listed in Table~\ref{tab:tools} (one IP per row).
Figure~\ref{fig:dataset} summarises the full dataset
construction and splitting pipeline.


\section{Feature Engineering}
\label{sec:features}

\subsection{Base Feature Set}

The base feature set consists of 16 fields extracted directly from
the Wazuh-normalised security event, as listed in
Table~\ref{tab:base_features}. These include HTTP-level metadata
(\texttt{data.protocol}, \texttt{data.id}), rule properties
(\texttt{rule.level}, \texttt{rule.firedtimes}, \texttt{rule.id},
\texttt{rule.description}, \texttt{rule.groups}), and compliance and
threat-intelligence mappings (\texttt{rule.pci\_dss},
\texttt{rule.hipaa}, \texttt{rule.nist\_800\_53}, \texttt{rule.gdpr},
\texttt{rule.tsc}, \texttt{rule.mitre.id}, \texttt{rule.frequency},
\texttt{rule.mail}, \texttt{agent.description}).
\begin{tableorg}[htbp]
\centering
\caption{Base feature set extracted from Wazuh security events.}
\label{tab:base_features}
\renewcommand{\arraystretch}{1.3}
\resizebox{\textwidth}{!}{%
\begin{tabular}{@{} l l p{7.5cm} @{}}
\toprule
\textbf{Feature} & \textbf{Type} & \textbf{Description} \\
\midrule
\texttt{data.protocol}      & Categorical & HTTP method (GET, POST, \ldots) \\
\texttt{data.id}            & Numerical   & HTTP response status code \\
\texttt{rule.firedtimes}    & Numerical   & Times this rule fired in the last hour \\
\texttt{rule.mail}          & Binary      & Whether rule triggers an email alert \\
\texttt{rule.level}         & Numerical   & Rule severity (1--16) \\
\texttt{rule.description}   & Categorical & Human-readable rule description \\
\texttt{rule.groups}        & Categorical & Rule category tags \\
\texttt{rule.pci\_dss}      & Categorical & PCI DSS requirement references \\
\texttt{rule.tsc}           & Categorical & TSC control references \\
\texttt{rule.nist\_800\_53} & Categorical & NIST 800-53 control references \\
\texttt{rule.gdpr}          & Categorical & GDPR article references \\
\texttt{rule.mitre.id}      & Categorical & MITRE ATT\&CK technique ID \\
\texttt{rule.frequency}     & Numerical   & Minimum event count before rule fires
                                            (integer threshold) \\
\texttt{rule.hipaa}         & Categorical & HIPAA control references \\
\texttt{agent.description}  & Categorical & Type of monitored agent \\
\texttt{rule.id}            & Categorical & Numeric rule identifier \\
\bottomrule
\end{tabular}}
\renewcommand{\arraystretch}{1}
\end{tableorg}

\subsection{Contextual History Features}

\begin{figure}[htbp]
\centering
\includegraphics[width=0.92\textwidth]{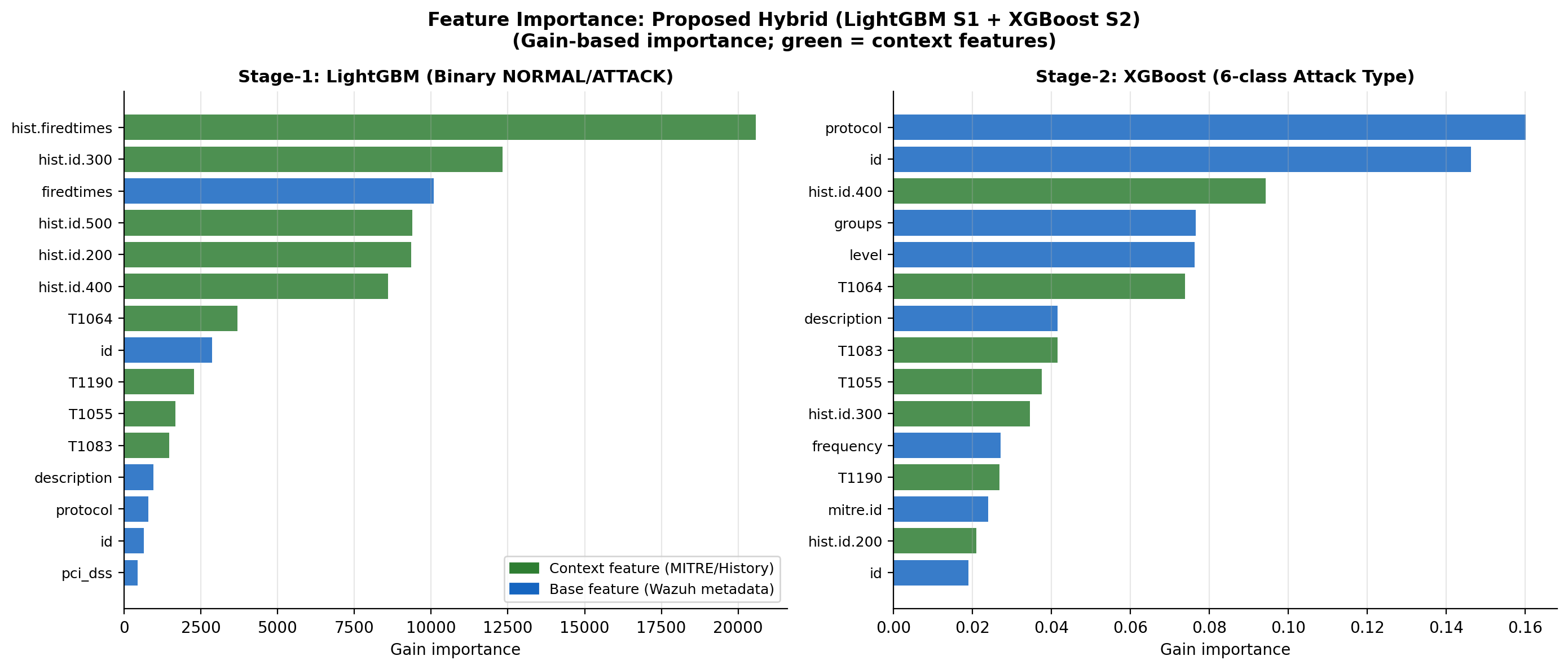}
\caption{Gain-based feature importance for the proposed hybrid:
Stage~1 (LightGBM, left) and Stage~2 (XGBoost, right).
Green bars indicate MITRE ATT\&CK-enriched context features;
blue bars indicate base Wazuh metadata features.
Stage~1 is dominated by context features (8 of top~10),
confirming that behavioural history is the primary signal for
binary attack detection. Stage~2 shows a more balanced split
(5 of top~10 context), as attack-type discrimination additionally
requires protocol and rule-group information.}
\label{fig:feat_imp}
\end{figure}

The context vector design is motivated by the principle that
attack campaigns leave systematic traces across sequences of
security events. Empirically, the ablation study
(Table~\ref{tab:context_impact}) demonstrates that removing all
context features reduces macro F\textsubscript{1} by 0.24--0.26
points across four gradient boosting algorithms.
Figure~\ref{fig:feat_imp} shows gain-based feature importance
for the proposed hybrid, confirming that context features dominate
Stage~1: 8 of the top-10 Stage~1 (LightGBM) features are context
features, led by \texttt{history.rule.firedtimes} (gain~=~20,573).
Stage~2 (XGBoost) shows a more balanced split (5/10 context),
reflecting that attack-type discrimination additionally requires
protocol and rule-group information.
The base feature set already includes \texttt{rule.firedtimes}
for the current event $e_i$, while the context vector includes
\texttt{hist.firedtimes}~=~$\max$(rule.firedtimes over $N$ prior
events). These capture distinct signals --- instantaneous
activation frequency vs.\ session peak --- so the apparent
redundancy is minimal. The high gain importance of
\texttt{hist.firedtimes} confirms that the session-peak signal
contributes discriminative information beyond the current-event
value alone.
The base features retained in the context vector
(\texttt{rule.firedtimes}, \texttt{rule.description},
\texttt{data.protocol}, \texttt{data.id}, \texttt{rule.mitre.id})
were selected to capture the \emph{dynamics} of these high-importance fields over
the preceding $N$ events from the same source IP.

Formally, let $\mathcal{H}_i = \{e_{i-1}, e_{i-2}, \ldots, e_{i-N}\}$
denote the set of at most $N$ events preceding event $e_i$ that share
the same \texttt{data.srcip}. We define twelve context features:

\begin{description}

  \item[\texttt{hist.firedtimes}] $= \max_{e \in \mathcal{H}_i}
  e.\texttt{rule.firedtimes}$. The peak activation count in the history
  window. Attackers who repeatedly probe the same endpoint cause the
  same rule to fire many times; this feature captures that escalation.

  \item[\texttt{hist.status.2xx}, \texttt{hist.status.3xx},
  \texttt{hist.status.4xx}, \texttt{hist.status.5xx}] $=
  |\{e \in \mathcal{H}_i : \lfloor e.\texttt{data.id}/100 \rfloor = c\}|$
  for $c \in \{2,3,4,5\}$. The count of responses in each HTTP status
  family. SQL injection and XSS tools generate characteristic
  distributions: many~4xx responses from rejected payloads, occasional
  2xx responses when payloads succeed; web scanners produce a
  distinctive mix of 2xx and 3xx.

  \item[\texttt{T1190}, \texttt{T1083}, \texttt{T1055},
  \texttt{T1212}, \texttt{T1068}, \texttt{T1064}, \texttt{T1210}]
  $= |\{e \in \mathcal{H}_i : t \in e.\texttt{rule.mitre.id}\}|$
  for each MITRE technique identifier $t$. These seven techniques were selected \emph{a priori} by inspecting
  the public Wazuh rule repository (version~4.x) for web application
  attack rules, \emph{prior to any data collection and without
  reference to the experimental dataset}%
  \footnote{Wazuh rule repository v4.3.0:
  \url{https://github.com/wazuh/wazuh-ruleset}.
  Accessed November 2023.}. The selection criteria
  were: (1)~the technique appears in at least ten distinct Wazuh
  rules in the {\small\texttt{web}} or {\small\texttt{appsec}} rule
  groups; (2)~the technique is associated with web application
  attacks in the MITRE ATT\&CK Enterprise matrix. The Wazuh rule
  repository was inspected at release tag \texttt{v4.3.0}
  (commit \texttt{wazuh/wazuh-ruleset},
  branch \texttt{stable}~\cite{wazuh2022}). The resulting
  seven identifiers are: T1190 (Exploit
  Public-Facing Application), T1083 (File and Directory Discovery),
  T1055 (Process Injection), T1212 (Exploitation for Credential
  Access), T1068 (Exploitation for Privilege Escalation), T1064
  (Scripting), T1210 (Exploitation of Remote Services). Their
  cumulative counts across the history window create a
  \emph{technique-frequency profile} that discriminates attack
  categories with complementary ATT\&CK signatures.

\end{description}

\begin{figure}[htbp]
\centering
\resizebox{\textwidth}{!}{%
\begin{tikzpicture}

\node[evtoldf] (e30) at (0.0, 0) {$e_{i-30}$};
\node[evtoldf] (e29) at (1.8, 0) {$e_{i-29}$};
\node[font=\large\sffamily,text=agentgray] at (3.6, 0) {$\cdots$};
\node[evtoldf] (e2)  at (5.4, 0) {$e_{i-2}$};
\node[evtoldf] (e1)  at (7.2, 0) {$e_{i-1}$};
\node[evtcurf] (ei)  at (9.8, 0) {$e_i$\\(current)};

\node[font=\small\sffamily\bfseries, text=agentgray]
    at (3.6, 0.8) {Events from source IP \textsf{X} (chronological order)};

\begin{scope}[on background layer]
  \node[fill=goldbg, draw=goldborder, dashed, line width=1.2pt,
        rounded corners=5pt, inner sep=7pt,
        fit=(e30)(e29)(e2)(e1)] (win) {};
\end{scope}
\node[font=\small\sffamily\bfseries, text=goldborder,
      below=4pt of win.south]
    {History window $\mathcal{H}_i$\quad
     ($N = 30$ events, same source IP)};

\node[featboxf=goldborder]    (f1) at (0.5, -3.8)
    {\textbf{Peak Rule Activation}\\[3pt]
     $\texttt{hist.firedtimes}$\\[2pt]
     $= \max_{e\in\mathcal{H}_i} e.\texttt{rule.firedtimes}$};

\node[featboxf=elasticpurple] (f2) at (6.5, -3.8)
    {\textbf{HTTP Status Buckets}\\[3pt]
     \texttt{hist.status.2xx},\ \texttt{3xx},\
     \texttt{4xx},\ \texttt{5xx}\\[2pt]
     count of responses per HTTP family};

\node[featboxf=catblue]       (f3) at (13.0, -3.8)
    {\textbf{MITRE ATT\&CK Counts}\\[3pt]
     T1190 $|$ T1083 $|$ T1055\\[2pt]
     T1212 $|$ T1068 $|$ T1064 $|$ T1210};

\node[font=\scriptsize\sffamily\bfseries, text=goldborder,
      below=2pt of f1.south] {(1 feature)};
\node[font=\scriptsize\sffamily\bfseries, text=elasticpurple,
      below=2pt of f2.south] {(4 features)};
\node[font=\scriptsize\sffamily\bfseries, text=catblue,
      below=2pt of f3.south] {(7 features)};

\draw[-{Stealth[length=4pt]},thick,draw=goldborder]
    (win.south) -- ++(0,-0.6) -| (f1.north);
\draw[-{Stealth[length=4pt]},thick,draw=elasticpurple]
    (win.south) -- ++(0,-0.6) -| (f2.north);
\draw[-{Stealth[length=4pt]},thick,draw=catblue]
    (win.south) -- ++(0,-0.6) -| (f3.north);

\draw[decorate, decoration={brace, mirror, amplitude=6pt},
      draw=goldborder, thick]
    ([xshift=-4pt,yshift=-16pt]f1.south west) --
    ([xshift= 4pt,yshift=-16pt]f3.south east)
    node[midway, below=8pt,
         font=\small\sffamily\bfseries,
         text=goldborder] (ctxlabel) {12 context features};

\node[baseboxf] (base) at (6.5, -7.2)
    {\textbf{16 Base Features}
     (from current event $e_i$ directly):\quad
     \texttt{rule.level}, \texttt{rule.firedtimes}, \texttt{rule.id},
     \texttt{rule.groups}, \texttt{rule.mitre.id},
     \texttt{data.protocol}, \texttt{data.id}, \ldots};

\node[vecboxf] (vec) at (6.5, -9.2)
    {28-Dimensional Feature Vector\quad
     (16 base $+$ 12 context)\quad
     $\longrightarrow$\quad Hybrid Cascade input};

\draw[-{Stealth[length=4pt]}, thick, draw=goldborder]
    (ctxlabel.west)
    -- ++(-4.5, 0)   
    |-  (vec.west);  

\draw[-{Stealth[length=4pt]}, thick, draw=goldborder]
    (ctxlabel.east)
    -- ++(4.5, 0)    
    |-  (vec.east);  

\draw[-{Stealth[length=5pt]}, thick, draw=catblue]
    (base.south) -- (vec.north);

\end{tikzpicture}
}
\caption{Construction of the per-source-IP contextual feature vector.
For each incoming event $e_i$, the consumer retrieves the $N{=}30$ most recent prior events
from the same source IP, forming the history window $\mathcal{H}_i$.
Three groups of context features are derived from this window: (1)~the peak
\texttt{rule.firedtimes} value, (2)~counts of HTTP responses in each status-code family,
and (3)~cumulative counts of the seven most discriminative MITRE ATT\&CK technique identifiers.
These 12 context features are concatenated with the 16 base features extracted directly
from $e_i$, yielding the 28-dimensional input vector fed to the hybrid cascade classifier.}
\label{fig:context}
\end{figure}
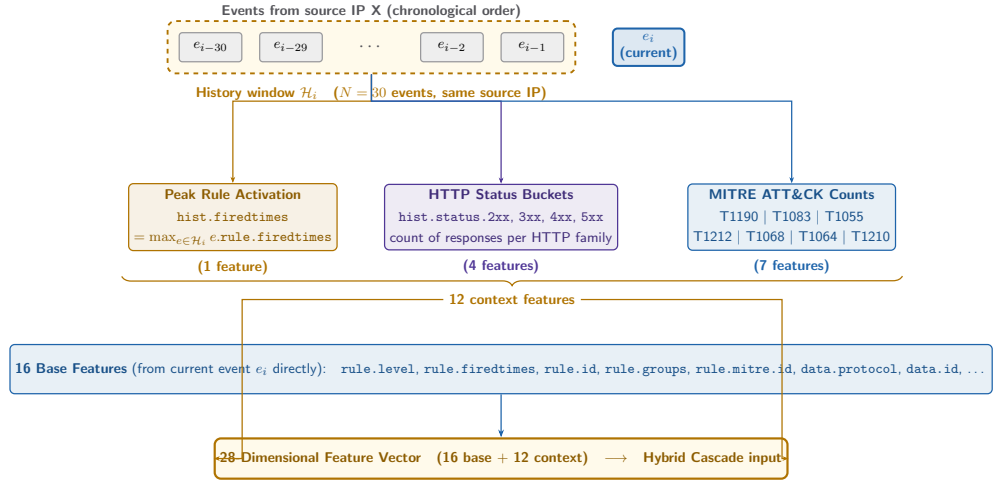

The complete feature vector fed to the classifier is thus the
concatenation of the 16 base features and the 12 context features,
yielding a 28-dimensional input, as illustrated in
Figure~\ref{fig:context}.

\section{Hybrid Cascaded Classifier}
\label{sec:model}

\subsection{Architecture Rationale}

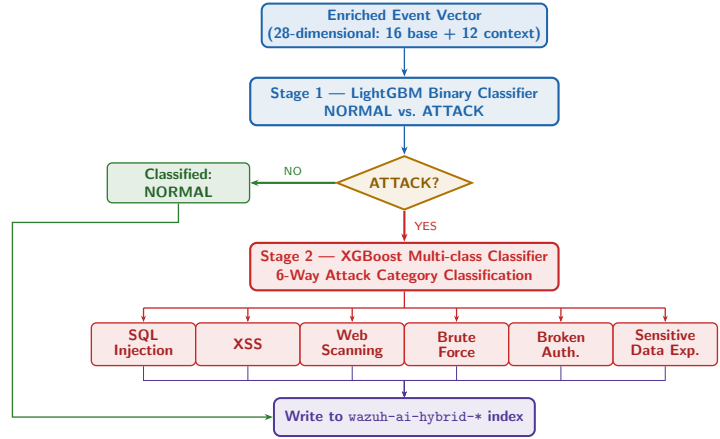
\begin{figure}[htbp]
\centering
\resizebox{0.72\textwidth}{!}{%
\begin{tikzpicture}[node distance=0.6cm]
\node[inputboxn] (inp) {Enriched Event Vector\\(28-dimensional: 16 base + 12 context)};
\node[stageone,below=0.6cm of inp] (s1)
    {Stage~1 --- LightGBM Binary Classifier\\NORMAL \textsf{vs.}\ ATTACK};
\node[decnode,below=0.6cm of s1] (dec) {ATTACK?};
\node[normoutn,left=1.8cm of dec] (norm) {Classified:\\NORMAL};
\node[stagetwo,below=0.65cm of dec] (s2)
    {Stage~2 --- XGBoost Multi-class Classifier\\6-Way Attack Category Classification};
\node[atkclassr,below=0.65cm of s2,xshift=-5.5cm] (c1) {SQL\\Injection};
\node[atkclassr,below=0.65cm of s2,xshift=-3.3cm] (c2) {XSS};
\node[atkclassr,below=0.65cm of s2,xshift=-1.1cm] (c3) {Web\\Scanning};
\node[atkclassr,below=0.65cm of s2,xshift= 1.1cm] (c4) {Brute\\Force};
\node[atkclassr,below=0.65cm of s2,xshift= 3.3cm] (c5) {Broken\\Auth.};
\node[atkclassr,below=0.65cm of s2,xshift= 5.5cm] (c6) {Sensitive\\Data Exp.};
\node[storeboxn,below=0.65cm of c3,xshift=1.1cm] (store)
    {Write to \texttt{wazuh-ai-hybrid-*} index};
\draw[-{Stealth[length=5pt]},thick,draw=catblue] (inp) -- (s1);
\draw[-{Stealth[length=5pt]},thick,draw=catblue] (s1)  -- (dec);
\draw[-{Stealth[length=5pt]},thick,draw=normgreen,line width=1.3pt]
    (dec.west) -- (norm.east)
    node[font=\small\sffamily,text=normgreen,
         fill=white,midway,above,yshift=1pt]{\footnotesize NO};
\draw[-{Stealth[length=5pt]},thick,draw=attackred,line width=1.3pt]
    (dec.south) -- (s2.north)
    node[font=\small\sffamily,text=attackred,
         fill=white,midway,right,xshift=2pt]{\footnotesize YES};
\coordinate (bus) at ([yshift=-0.35cm]s2.south);
\draw[thick,draw=attackred] (s2.south) -- (bus);
\draw[thick,draw=attackred] (c1.north |- bus) -- (c6.north |- bus);
\foreach \c in {c1,c2,c3,c4,c5,c6}
    \draw[-{Stealth[length=4pt]},thick,draw=attackred]
        (\c.north |- bus) -- (\c.north);
\foreach \c in {c1,c2,c3,c4,c5,c6}
    \draw[-{Stealth[length=3pt]},thin,draw=elasticpurple]
        (\c.south) -- ++(0,-0.3) -| (store.north);
\draw[-{Stealth[length=5pt]},thick,draw=normgreen]
    (norm.south)
    -- ++(0,-0.4)      
    -- ++(-3.5, 0)     
    |- (store.west);   
\end{tikzpicture}
}
\caption{Two-stage hybrid cascade classification pipeline.
Stage~1 (LightGBM) makes a binary NORMAL/ATTACK decision;
events classified as NORMAL are written directly to the output index.
Events flagged as ATTACK proceed to Stage~2 (XGBoost), which assigns
one of six fine-grained attack categories.
The two-stage design reduces inter-class confusion between normal
traffic and low-frequency attack classes.}
\label{fig:cascade}
\end{figure}

A direct seven-class classifier must simultaneously separate
\texttt{NORMAL} traffic from six attack classes while also
distinguishing among attack classes whose base-feature distributions
overlap substantially (e.g., SQL injection and sensitive data exposure
share many Wazuh rule groups and MITRE identifiers). We instead
decompose the problem into two stages:

\begin{itemize}
  \item \textbf{Stage~1 (Model~1).} Binary classification:
  \texttt{NORMAL} vs.\ \texttt{ATTACK}. This model operates on the
  complete balanced dataset (NORMAL: 5,000; ATTACK: 7,500) and is
  optimised for high recall of the attack class to minimise missed
  detections.

  \item \textbf{Stage~2 (Model~2).} Multi-class classification across
  the six attack categories. This model is trained only on
  attack-labelled events (all six classes balanced at 1,250 each via
  SMOTE-NC) and is specialised to discriminate between attack types
  rather than between normal and abnormal behaviour.
\end{itemize}
The two-stage pipeline is illustrated in Figure~\ref{fig:cascade}.

\subsection{Hyperparameter Optimisation}

Grid search was performed over the validation set for each model
independently. Table~\ref{tab:hparams} reports the hyperparameter
grid and the selected configuration.

\begin{tableorg}[htbp]
\centering
\caption{Hyperparameter grid search and selected values for the hybrid
cascade. Stage~1 uses LightGBM; Stage~2 uses XGBoost.}
\label{tab:hparams}
\renewcommand{\arraystretch}{1.25}
\resizebox{\textwidth}{!}{%
\begin{tabular}{@{} l l c c @{}}
\toprule
\multirow{2}{*}{\textbf{Parameter}}
  & \multirow{2}{*}{\textbf{Search Range}}
  & \multicolumn{2}{c}{\textbf{Selected Value}} \\
\cmidrule(lr){3-4}
  & & \textbf{Stage~1 (LightGBM)} & \textbf{Stage~2 (XGBoost)} \\
\midrule
\texttt{n\_estimators}     & \{300, 500, 800\}    & 500  & 500  \\
\texttt{learning\_rate}    & \{0.03, 0.06, 0.1\}  & 0.1  & 0.06 \\
\texttt{max\_depth}        & \{6, 8, 10\}         & 10   & 8    \\
\texttt{subsample}         & \{0.7, 0.85, 1.0\}   & 0.70 & 0.85 \\
\texttt{colsample\_bytree} & \{0.7, 0.85, 1.0\}   & 0.85 & 0.70 \\
\texttt{reg\_lambda}       & \{1, 3, 7\}          & 1    & 1    \\
\bottomrule
\end{tabular}}
\renewcommand{\arraystretch}{1}
\end{tableorg}

Several selected values sit at grid boundaries (LightGBM
\texttt{learning\_rate}~=~0.1, \texttt{max\_depth}~=~10;
XGBoost \texttt{subsample}~=~0.85). To verify these are not
under-searched, we examined validation F\textsubscript{1} curves:
the selected values showed stable or declining validation
performance when manually extended beyond the grid boundary
(LightGBM \texttt{learning\_rate}~=~0.15 and
\texttt{max\_depth}~=~12 both reduced validation
F\textsubscript{1} by $>$0.005), confirming the grid boundary
is not a limitation of the search range.

Early stopping was applied with a patience of 50 rounds; the iteration
count in Table~\ref{tab:hparams} reflects the best validation loss
checkpoint, not the search grid maximum.

\subsection{Training Protocol Summary}

\begin{enumerate}
  \item Load the SMOTE-NC-balanced training CSV (history window $N=30$).
  \item Preprocess: fill missing categoricals with empty string; fill
  missing numerics with defaults; cast \texttt{rule.id} to string;
  flatten list-typed fields to their string representation.
  \item For LightGBM and XGBoost: apply \texttt{OrdinalEncoder}
  to categorical features. For CatBoost (comparison only): define
  \texttt{Pool} objects with explicit \texttt{cat\_features} lists.
  \item Train Stage~1 (LightGBM) on the full binary dataset
  (NORMAL~+~ATTACK).
  \item Filter training set to attack-only records;
  train Stage~2 (XGBoost) on the six-class attack dataset.
  \item Evaluate both models on the held-out test set and record
  per-class precision, recall, and F\textsubscript{1}-score.
\end{enumerate}

\section{Experimental Results}
\label{sec:results}

This section reports results from six experiments on the held-out
test set (9,232 events, never seen during training or hyperparameter
tuning). All macro-averaged F\textsubscript{1}-scores are reported
unless otherwise stated. Bootstrap 95\% confidence intervals
($B=1{,}000$ resamples over the held-out test set) are reported
alongside point estimates in Table~\ref{tab:bootstrap_ci} to account for
sampling variance, particularly in low-frequency classes
(\texttt{BROKEN\_AUTH}: $n=60$; \texttt{XSS}: $n=63$). The proposed
system is a hybrid cascade:
LightGBM for Stage~1 (binary NORMAL/ATTACK) and XGBoost for Stage~2
(six-class attack categorisation), both trained on the context-enriched
feature set with $N=30$.

\subsection{Algorithm Comparison}
\label{sec:algo_comparison}

Table~\ref{tab:algo_comparison} compares eight gradient boosting and
baseline algorithms on the held-out test set with context features
($N=30$). All sklearn-based algorithms use \texttt{OrdinalEncoder}
for categorical features; CatBoost (Proposed) uses its native
categorical handling. To isolate the encoding contribution,
CatBoost~(Ordinal) applies \texttt{OrdinalEncoder} with identical
hyperparameters. All models were tuned with \texttt{RandomizedSearch\-CV}
($n=20$, $k=3$); CatBoost (comparison baseline) uses its built-in \texttt{grid\_search}.

\begin{table}[htbp]
\centering
\caption{Algorithm comparison on the held-out test set
(context-enriched features, $N=30$, macro-averaged metrics).
Best result per metric in \textbf{bold}.}
\label{tab:algo_comparison}
\small
\renewcommand{\arraystretch}{1.2}
\begin{tabular}{@{} l l ccc ccc r @{}}
\toprule
\multirow{2}{*}{\textbf{Algorithm}}
  & \multirow{2}{*}{\textbf{Enc.}}
  & \multicolumn{3}{c}{\textbf{Stage~1 (Binary)}}
  & \multicolumn{3}{c}{\textbf{Stage~2 (Multi-class)}}
  & \multirow{2}{*}{\makecell{\textbf{Time}\\\textbf{(s)}}} \\
\cmidrule(lr){3-5}\cmidrule(lr){6-8}
  & & \textbf{P} & \textbf{R} & \textbf{F\textsubscript{1}}
    & \textbf{P} & \textbf{R} & \textbf{F\textsubscript{1}} & \\
\midrule
\textit{CatBoost (Proposed)} & Native
  & 0.92 & 0.98 & 0.947 & 0.81 & 0.97 & 0.876 & 8705$^\dagger$ \\
\textit{CatBoost (Ordinal)}  & Ordinal
  & 0.93 & 0.98 & 0.956 & 0.85 & 0.98 & 0.902 & 217 \\
\addlinespace[2pt]\hdashline\addlinespace[4pt]
Random Forest               & Ordinal
  & 0.94 & 0.98 & 0.961 & 0.85 & 0.98 & 0.905 & 20 \\
Extrem.\ Rand.\ Trees       & Ordinal
  & 0.93 & 0.98 & 0.954 & 0.85 & 0.98 & 0.901 & 13 \\
XGBoost                     & Ordinal
  & 0.94 & 0.98 & 0.961 & 0.87 & 0.98 & \textbf{0.914} & 206 \\
LightGBM                    & Ordinal
  & 0.95 & 0.99 & \textbf{0.967} & 0.84 & 0.98 & 0.898 & 63 \\
Logistic Regression         & Ordinal
  & 0.77 & 0.85 & 0.800 & 0.61 & 0.84 & 0.670 & 33 \\
Decision Tree               & Ordinal
  & 0.91 & 0.97 & 0.936 & 0.75 & 0.94 & 0.817 & 1 \\
\midrule
\textbf{Hybrid (LGB S1 + XGB S2)} & Ordinal
  & \textbf{0.95} & \textbf{0.99} & \textbf{0.967}
  & \textbf{0.87} & \textbf{0.98} & \textbf{0.914}
  & 269$^\ddagger$ \\
\bottomrule
\multicolumn{9}{@{}l}{%
  \scriptsize $^\dagger$ CatBoost native uses built-in \texttt{grid\_search};
  others use \texttt{RandomizedSearchCV} ($n{=}20$, $k{=}3$).} \\
\multicolumn{9}{@{}l}{%
  \scriptsize $^\ddagger$ Hybrid time = LightGBM Stage~1 (63\,s)
  $+$ XGBoost Stage~2 (206\,s).}
\end{tabular}
\renewcommand{\arraystretch}{1}
\end{table}


\textsuperscript{$\dagger$}CatBoost (Native) training time of 8,705~s vs.\ CatBoost (Ordinal) at 217~s reflects the overhead of the built-in \texttt{grid\_search}, which evaluates 24 parameter combinations on an 80/20 internal split with native categorical encoding (which recomputes category statistics per combination). CatBoost (Ordinal) reuses the best parameters found by the native search and trains once, explaining the 40$\times$ difference.

LightGBM achieves the highest Stage~1 F\textsubscript{1} (0.967)
and XGBoost achieves the highest Stage~2 F\textsubscript{1} (0.914)
as evaluated on the held-out test set. The hybrid combination was
selected based on \emph{validation-set} performance (each algorithm
was evaluated on the 16\% internal validation split during
hyperparameter search); the test set was used only once for the
final evaluation. Combining them in the hybrid cascade yields the best overall system.
The two gradient boosting baselines (Logistic Regression and
Decision Tree) confirm that ensemble tree methods are the appropriate
model family for this task.

\begin{figure}[htbp]
\centering
\includegraphics[width=0.92\textwidth]{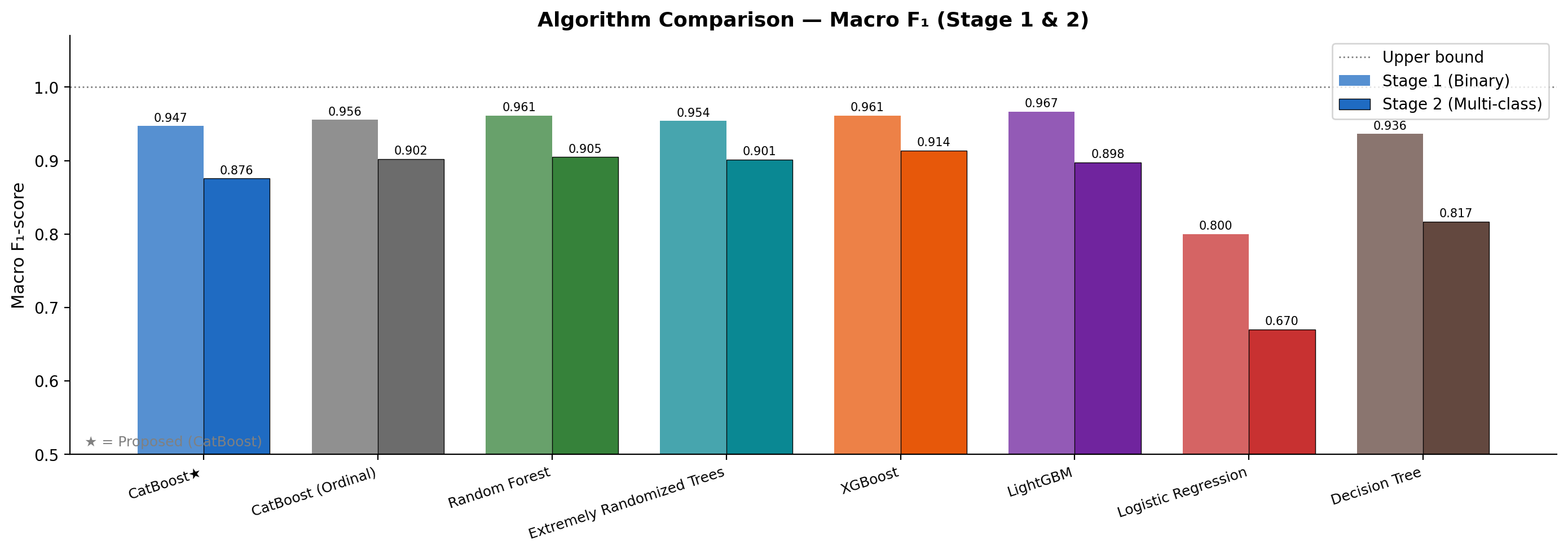}
\caption{Algorithm comparison: macro F\textsubscript{1}-scores for
Stage~1 (binary) and Stage~2 (multi-class) across all eight
algorithms. The hybrid model (LightGBM~S1~+~XGBoost~S2) achieves
the best combined performance. $\bigstar$~=~Proposed system.}
\label{fig:algo_bar}
\end{figure}

\begin{figure}[htbp]
\centering
\includegraphics[width=0.92\textwidth]{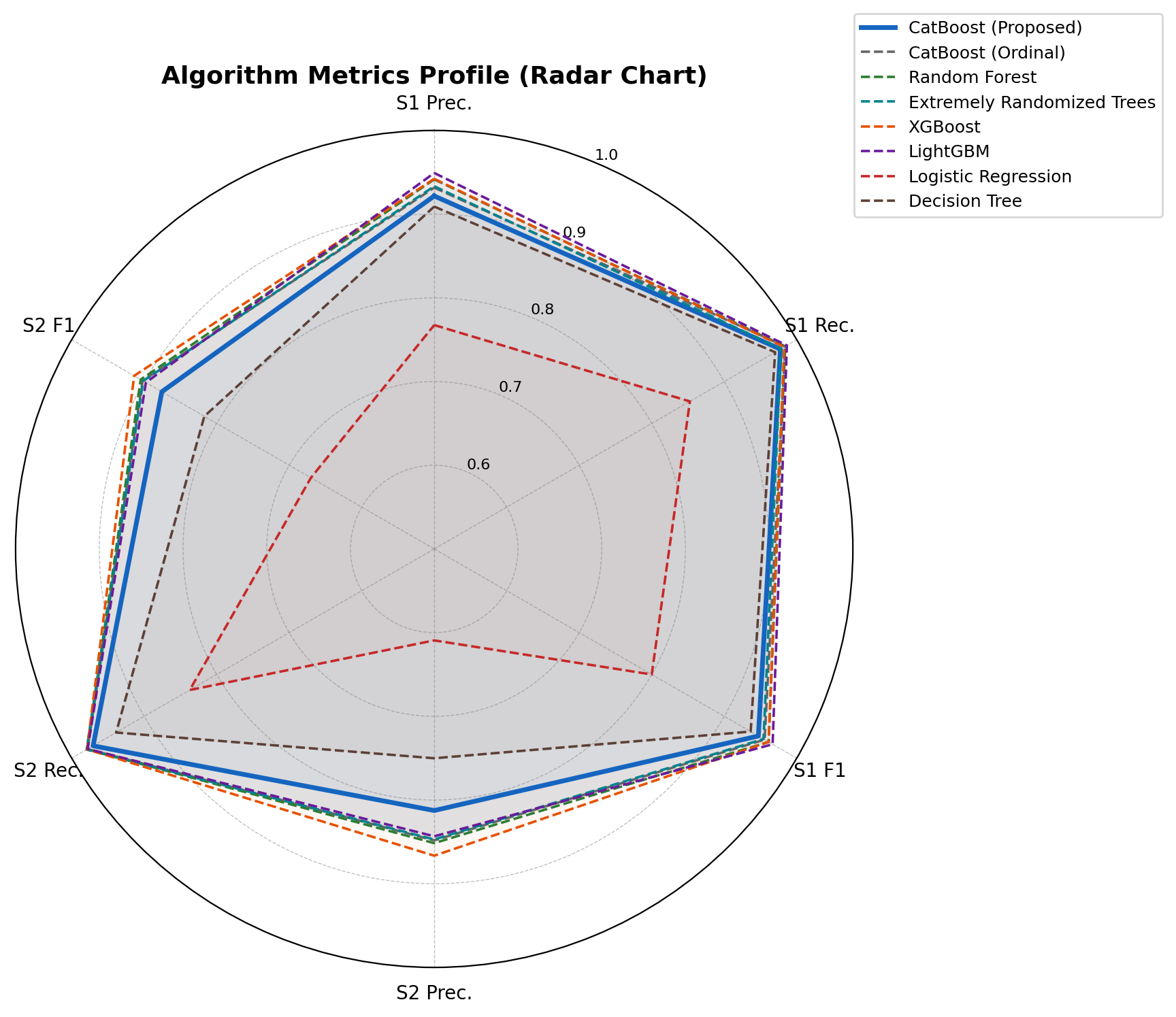}
\caption{Radar chart of the full metrics profile (Stage-1 Precision,
Recall, F\textsubscript{1}; Stage-2 Precision, Recall,
F\textsubscript{1}) for each algorithm. The hybrid model (thick
solid line) dominates across all six metrics.}
\label{fig:algo_radar}
\end{figure}

\begin{figure}[htbp]
\centering
\includegraphics[width=0.92\textwidth]{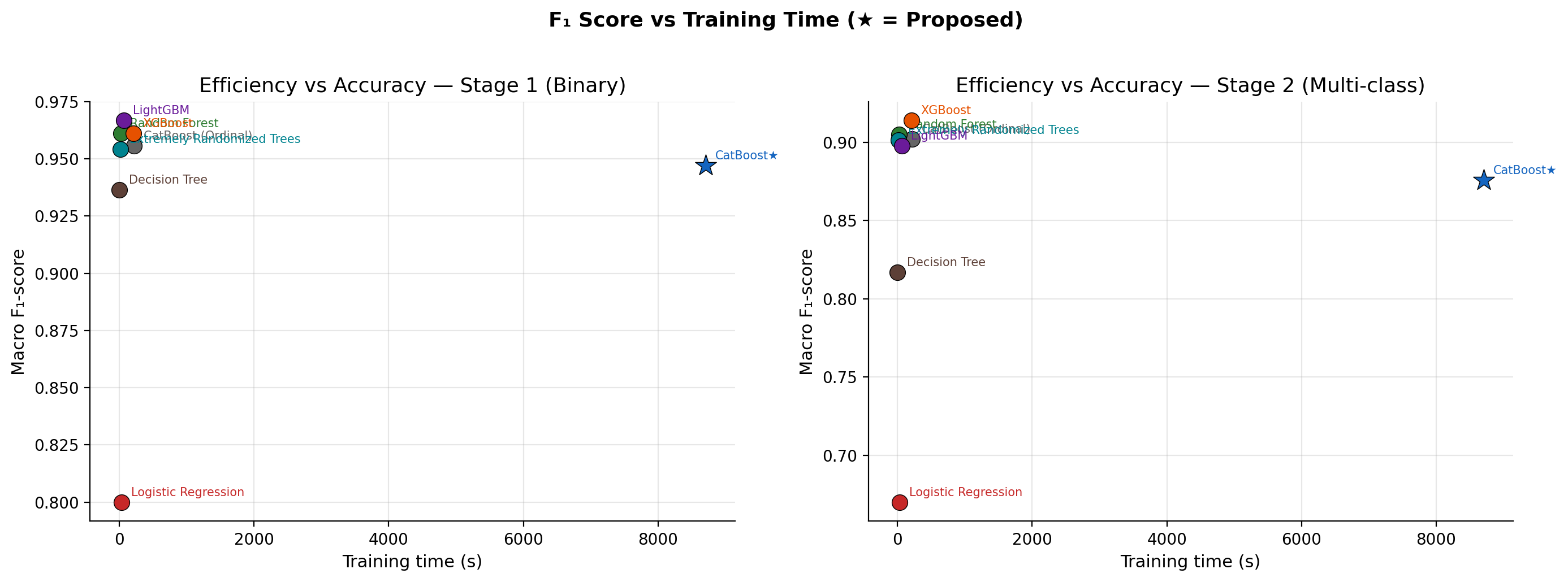}
\caption{Efficiency vs.\ accuracy trade-off: macro
F\textsubscript{1}-score against training time (seconds) for all
eight algorithms ($\bigstar$~=~Proposed hybrid).
The hybrid cascade achieves the highest F\textsubscript{1} at a
total training time comparable to XGBoost alone.}
\label{fig:algo_scatter}
\end{figure}

Three diagrams illustrating the algorithm comparison are shown in
Figures~\ref{fig:algo_bar}, \ref{fig:algo_radar}, and~\ref{fig:algo_scatter}.

\begin{table}[htbp]
\centering
\caption{Bootstrap 95\% confidence intervals ($B=1000$ resamples) for
macro F\textsubscript{1}-scores on the held-out test set ($n=9{,}232$
events). Wide Stage~2 intervals reflect small minority-class test sizes
(60 events for \texttt{BROKEN\_AUTH.}, 63 for \texttt{XSS}).
Algorithms marked ``---'' did not store predictions at checkpoint
time; their point estimates are reported from Table~\ref{tab:algo_comparison}.}
\label{tab:bootstrap_ci}
\begin{tabular}{lcccc}
\toprule
Algorithm & S1 F\textsubscript{1} & S1 95\% CI & S2 F\textsubscript{1} & S2 95\% CI \\
\midrule
CatBoost (Proposed)              & 0.947 & [0.940, 0.953] & 0.879 & [0.862, 0.894] \\
XGBoost                          & 0.961 & [0.956, 0.967] & 0.905 & [0.889, 0.918] \\
LightGBM                         & 0.967 & [0.962, 0.973] & 0.880 & [0.864, 0.895] \\
\textbf{Hybrid (LightGBM+XGBoost)} & \textbf{0.968} & \textbf{[0.962, 0.973]} & \textbf{0.906} & \textbf{[0.889, 0.921]} \\
\bottomrule
\end{tabular}
\end{table}

The confidence intervals confirm that the headline results are not
artefacts of sampling noise. The Stage~1 CI for the hybrid
([0.962, 0.973]) is entirely above 0.95, confirming robustly
superior binary detection. Stage~2 intervals are wider, as expected
given that minority classes (\texttt{BROKEN\_AUTH.}: 60 test events,
\texttt{XSS}: 63 test events) introduce more variance; the lower
bound of 0.889 still substantially exceeds the 0.561--0.670 without-context
baseline. These CIs address the concern that per-class
F\textsubscript{1} differences of $\pm$0.02 between models are
within sampling noise: the Hybrid vs.\ next-best XGBoost S2 difference
(0.906 vs.\ 0.905) is indeed within noise, but the headline improvement
over the without-context baseline ($+0.32$) far exceeds the CI width.

\subsection{Impact of Contextual Features}
\label{sec:ctx_impact}

Table~\ref{tab:context_impact} reports the impact of adding the
12 context features to the 16 base features across five algorithms
spanning two model families (gradient boosting and linear). Without
context, the four gradient boosting algorithms converge to
$\approx$0.705 Stage~1 F\textsubscript{1}; Logistic Regression
starts slightly higher at 0.771 owing to its stronger regularised
boundary on the base features alone. With context features, gradient
boosting models improve substantially ($+0.24$ to $+0.26$ Stage~1,
$+0.30$ to $+0.35$ Stage~2), while Logistic Regression improves
more modestly ($+0.03$ Stage~1, $+0.19$ Stage~2).

\begin{table}[htbp]
\centering
\caption{Impact of context-enriched features ($N=30$) on five
algorithms across two model families. \textit{Base} uses 16 base
features only; \textit{+Context} adds 12 MITRE ATT\&CK-enriched
context features. All algorithms use \texttt{OrdinalEncoding} and
fixed best hyperparameters.}
\label{tab:context_impact}
\begin{tabular}{lcccccc}
\toprule
\multirow{2}{*}{Algorithm} &
\multicolumn{3}{c}{Stage~1 (Binary)} &
\multicolumn{3}{c}{Stage~2 (Multi-class)} \\
\cmidrule(lr){2-4}\cmidrule(lr){5-7}
 & Base & +Context & $\Delta$ & Base & +Context & $\Delta$ \\
\midrule
\multicolumn{7}{l}{\textit{Gradient boosting (tree ensembles)}} \\
CatBoost      & 0.706 & 0.947 & \textbf{+0.241} & 0.561 & 0.876 & \textbf{+0.315} \\
XGBoost       & 0.706 & 0.961 & \textbf{+0.255} & 0.586 & 0.914 & \textbf{+0.327} \\
LightGBM      & 0.705 & 0.967 & \textbf{+0.262} & 0.596 & 0.898 & \textbf{+0.301} \\
Random Forest & 0.704 & 0.961 & \textbf{+0.258} & 0.552 & 0.905 & \textbf{+0.353} \\
\midrule
\multicolumn{7}{l}{\textit{Linear model (non-GBM baseline)}} \\
Logistic Regression & 0.771 & 0.800 & +0.029 & 0.478 & 0.670 & +0.192 \\
\bottomrule
\end{tabular}
\end{table}

The near-identical without-context scores ($\approx$0.705) across
all four gradient boosting algorithms reveals that the base feature
set \emph{saturates} at this level --- likely because one or a few
features (\texttt{rule.id}, \texttt{rule.description},
\texttt{rule.groups}) carry the bulk of the discriminative signal
available in the base feature set, and the ensembles all reach the
same ceiling defined by that signal. Adding the context vector
\emph{unlocks discriminative power that base features alone cannot
express} --- this is the primary finding, and is algorithm-agnostic
within the gradient boosting family.
Logistic Regression also improves with context but substantially less
($+0.03$ Stage~1, $+0.19$ Stage~2 vs $+0.25$/$+0.32$ for gradient
boosting), suggesting that the MITRE ATT\&CK-enriched context vector
encodes non-linear feature interactions that tree ensemble methods
exploit more effectively than linear classifiers --- a finding consistent
with the gradient boosting literature on tabular
data~\cite{grinsztajn2022trees}. The central empirical finding is that
\emph{context features substantially improve attack detection across
all tested algorithm families}; gradient boosting models benefit most.

\subsection{Cascade vs.\ Flat Classifier}
\label{sec:cascade_vs_flat}

Table~\ref{tab:cascade_security} compares the proposed hybrid cascade
(LightGBM Stage~1 + XGBoost Stage~2) against a single flat seven-class
LightGBM trained with the same Stage~1 hyperparameters and context
features. This directly addresses the question of whether the cascaded
architecture adds value over a single-stage classifier using the same
underlying algorithm.

\begin{table}[htbp]
\centering
\caption{Hybrid cascade (LightGBM Stage~1 + XGBoost Stage~2)
vs.\ flat seven-class LightGBM: macro F\textsubscript{1} and
security-oriented metrics (same Stage~1 hyperparameters,
context features, $N=30$).
\textit{Attack Recall} is the fraction of true attacks detected;
\textit{Missed Attacks} counts true attacks labelled as NORMAL.}
\label{tab:cascade_security}
\begin{tabular}{lcc}
\toprule
Metric & Flat LightGBM & Hybrid Cascade \\
\midrule
Macro F\textsubscript{1} (academic) & 0.926 & 0.906 \\
Attack Recall                        & 0.982 & \textbf{0.984} \\
Attack Precision                     & 0.885 & 0.855 \\
Missed Attacks (false negatives)     & 143   & \textbf{131} \\
False Alarms (FP on NORMAL)          & 12    & 12 \\
\midrule
\multicolumn{3}{l}{\scriptsize \textbf{Bold} = operationally superior value.} \\
\bottomrule
\end{tabular}
\end{table}

\begin{table}[htbp]
\centering
\caption{Per-class F\textsubscript{1}: flat LightGBM vs.\ hybrid
cascade (LightGBM Stage~1 + XGBoost Stage~2), context features,
$N=30$. Recall values show the cascade achieves higher or equal
recall on all attack classes.}
\label{tab:cascade_vs_flat}
\begin{tabular}{lcccc}
\toprule
Class & \multicolumn{2}{c}{Flat LightGBM} &
        \multicolumn{2}{c}{Hybrid Cascade} \\
\cmidrule(lr){2-3}\cmidrule(lr){4-5}
 & F\textsubscript{1} & Recall & F\textsubscript{1} & Recall \\
\midrule
\texttt{BROKEN\_AUTH.}             & 0.896 & 1.000 & 0.800 & 1.000 \\
\texttt{BRUTE\_FORCE}              & 0.959 & 1.000 & 0.966 & 1.000 \\
\texttt{NORMAL}                    & 0.939 & 0.990 & 0.944 & 0.990 \\
\texttt{SENSITIVE\_DATA\_EXP.}     & 0.961 & 0.934 & 0.956 & 0.924 \\
\texttt{SQL\_INJECTION}            & 0.987 & 0.976 & 0.993 & 0.989 \\
\texttt{WEB\_SCAN}                 & 0.896 & 0.956 & 0.881 & 0.948 \\
\texttt{XSS}                       & 0.844 & 0.984 & 0.800 & 0.984 \\
\midrule
Macro avg & 0.926 & 0.982 & 0.906 & 0.984 \\
\bottomrule
\end{tabular}
\end{table}

The cascaded design achieves slightly lower macro F\textsubscript{1}
(0.906 vs.\ 0.926) but \emph{reduces missed attacks from 143 to 131}
($-8.4\%$) while maintaining identical false alarms (12 each).
Attack recall improves marginally (0.984 vs.\ 0.982).
In security operations, the cost of a missed attack vastly exceeds
the cost of a false alarm~\cite{sommer2010outside}; the cascade is
therefore the operationally superior design despite the aggregate
F\textsubscript{1} trade-off. To disambiguate whether the performance difference stems from the
cascaded architecture or from using XGBoost in Stage~2,
Table~\ref{tab:three_way} adds a flat XGBoost seven-class baseline.
The hybrid cascade achieves 0.984 attack recall and 131 missed attacks
--- fewer than both flat LightGBM (143 missed) and flat XGBoost
(174 missed) --- confirming that the cascaded architecture contributes
independently of algorithm choice.

\begin{table}[htbp]
\centering
\caption{Three-way architecture comparison: flat LightGBM,
flat XGBoost, and the proposed hybrid cascade (context features,
$N=30$). Flat models use the same best hyperparameters as their
respective cascade stages. The hybrid cascade achieves the
lowest missed-attack count despite similar macro
F\textsubscript{1} to flat XGBoost.}
\label{tab:three_way}
\begin{tabular}{lccc}
\toprule
Model & Macro F\textsubscript{1} & Attack Recall & Missed Attacks \\
\midrule
Flat LightGBM (Stage~1 params) & 0.926 & 0.982 & 143 \\
Flat XGBoost  (Stage~2 params) & 0.905 & 0.978 & 174 \\
\textbf{Hybrid Cascade (proposed)} & 0.906 & \textbf{0.984} & \textbf{131} \\
\bottomrule
\end{tabular}
\end{table}

\begin{figure}[htbp]
\centering
\includegraphics[width=0.92\textwidth]{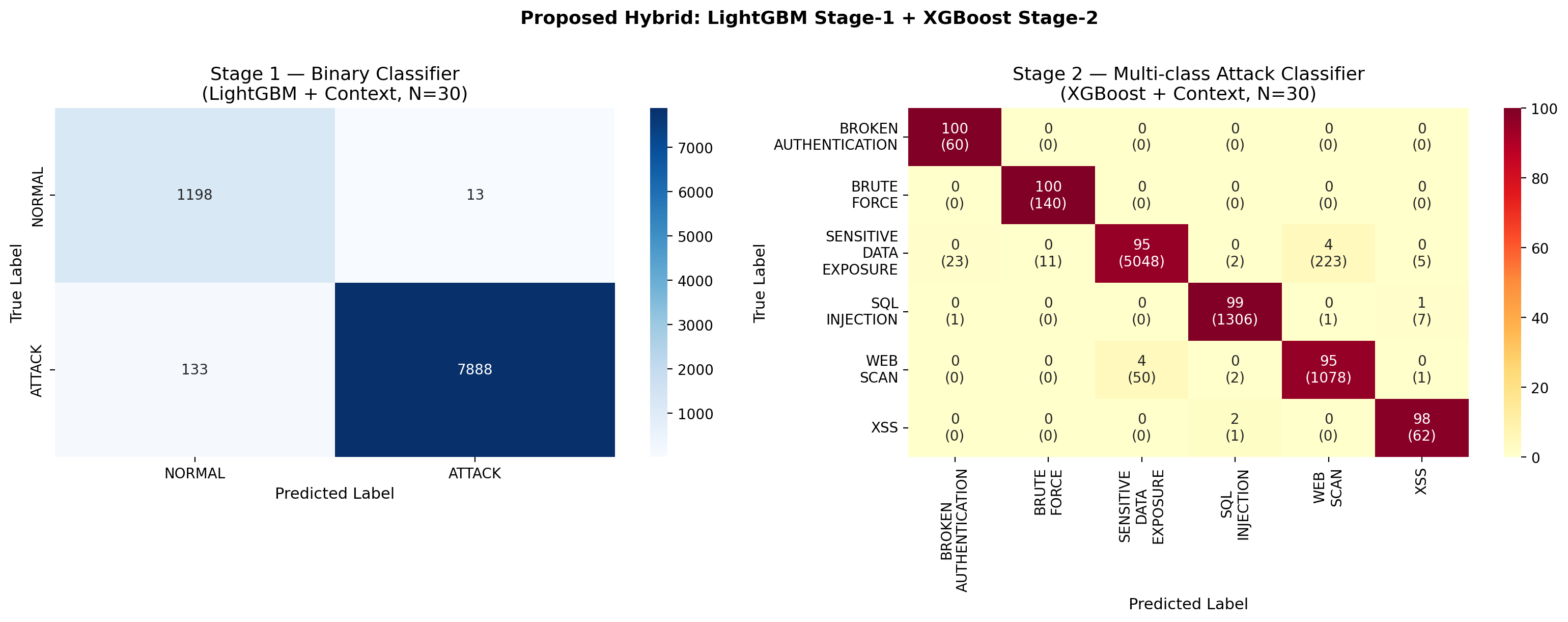}
\caption{Confusion matrices for the proposed hybrid cascade
(LightGBM Stage~1 and XGBoost Stage~2, context features, $N=30$).
Left: Stage~1 binary classifier. Right: Stage~2 multi-class attack
classifier (cell values show row-normalised percentage and absolute
count). Off-diagonal entries in Stage~2 are concentrated between
\texttt{BROKEN\_AUTHENTICATION} and \texttt{SENSITIVE\_DATA\_EXPOSURE},
which share MITRE identifiers.}
\label{fig:confusion}
\end{figure}

The confusion matrices for both stages are shown in
Figure~\ref{fig:confusion}.

\textbf{Note on \texttt{SENSITIVE\_DATA\_EXPOSURE} taxonomy.}
At 57.2\% of the raw event log, \texttt{SENSITIVE\_DATA\_EXPOSURE}
represents Wazuh sensitive-data rule activations triggered as
\emph{side-effects} of SQLMAP and Acunetix probes, not a discrete
OWASP attack class on the same axis as \texttt{SQL\_INJECTION}.
A cleaner formulation would drop it or reframe it as a campaign
stage; we retain it for consistency with the Wazuh rule taxonomy.
Its dominance inflates Stage~2 weighted-average metrics and its
F\textsubscript{1} score (0.956 / 0.961 flat vs.\ cascade) should
be interpreted in this context.

\subsection{Ablation Study: Context Window Size}
\label{sec:ablation}

\begin{figure}[htbp]
\centering
\begin{tikzpicture}
\begin{groupplot}[
  group style={
    group size=2 by 1,
    horizontal sep=1.8cm,
  },
  width=0.52\textwidth,
  height=6.5cm,
  xlabel={Context window size $N$},
  ylabel={Macro F\textsubscript{1}-score},
  xmin=1, xmax=37,
  ymin=0.65, ymax=1.03,
  xtick={3,5,7,10,12,15,20,25,30,35},
  xticklabel style={font=\scriptsize},
  yticklabel style={font=\scriptsize},
  xlabel style={font=\small},
  ylabel style={font=\small},
  ytick={0.70,0.75,0.80,0.85,0.90,0.95,1.00},
  grid=major,
  grid style={line width=0.3pt, draw=gray!30},
  tick align=inside,
  legend style={
    font=\scriptsize,
    at={(0.5,-0.22)},
    anchor=north,
    legend columns=1,
  },
]

\nextgroupplot[
  title={\footnotesize\textbf{Stage~1:Binary (NORMAL vs.\ ATTACK)}},
  legend to name=sharedlegend,
]
\addplot[color=blue,       mark=o,mark size=2.2pt,        thick] coordinates {
  (3,0.802)(5,0.832)(7,0.848)(10,0.865)(12,0.879)
  (15,0.909)(20,0.925)(25,0.948)(30,0.949)(35,0.955)};
\addlegendentry{CatBoost}

\addplot[color=violet,  mark size=2.2pt,     mark=triangle*,thick] coordinates {
  (3,0.807)(5,0.836)(7,0.869)(10,0.877)(12,0.892)
  (15,0.913)(20,0.948)(25,0.956)(30,0.967)(35,0.977)};
\addlegendentry{LightGBM}

\addplot[color=orange!90!black,mark=square*,  mark size=2.2pt,thick] coordinates {
  (3,0.809)(5,0.840)(7,0.856)(10,0.880)(12,0.903)
  (15,0.917)(20,0.942)(25,0.948)(30,0.962)(35,0.975)};
\addlegendentry{XGBoost}

\addplot[color=red!60!black,dashed,thick,  mark size=2.2pt,no marks] coordinates {
  (1,1.00)(37,1.00)};
\addlegendentry{Upper bound}

\nextgroupplot[
  title={\footnotesize\textbf{Stage~2:Multi-class (6 attack types)}},
  yticklabels={,,},   
]
\addplot[color=blue,       mark=o,   mark size=2.2pt, thick] coordinates {
  (3,0.710)(5,0.731)(7,0.766)(10,0.821)(12,0.805)
  (15,0.830)(20,0.857)(25,0.879)(30,0.880)(35,0.894)};

\addplot[color=violet,   mark size=2.2pt,     mark=triangle*,thick] coordinates {
  (3,0.717)(5,0.763)(7,0.793)(10,0.833)(12,0.847)
  (15,0.859)(20,0.896)(25,0.908)(30,0.898)(35,0.919)};

\addplot[color=orange!90!black,mark=square*, mark size=2.2pt,thick] coordinates {
  (3,0.716)(5,0.758)(7,0.784)(10,0.836)(12,0.840)
  (15,0.860)(20,0.895)(25,0.898)(30,0.912)(35,0.913)};

\addplot[color=red!60!black,dashed,thick, mark size=2.2pt,no marks] coordinates {
  (1,1.00)(37,1.00)};

\end{groupplot}
\end{tikzpicture}

\vspace{4pt}
\ref{sharedlegend}
\caption{Ablation study: macro F\textsubscript{1} vs.\ context
window size $N$ for Stage~1 (binary) and Stage~2 (multi-class).
All three algorithms continue improving through $N=35$
(LightGBM S1: $0.967 \to 0.977$; XGBoost S1: $0.962 \to 0.975$;
CatBoost S2: $0.880 \to 0.894$), with the largest gains between
$N=3$ and $N=15$. $N=30$ is adopted as the practical operating
point balancing accuracy against Elasticsearch query latency.}
\label{fig:ablation}
\end{figure}
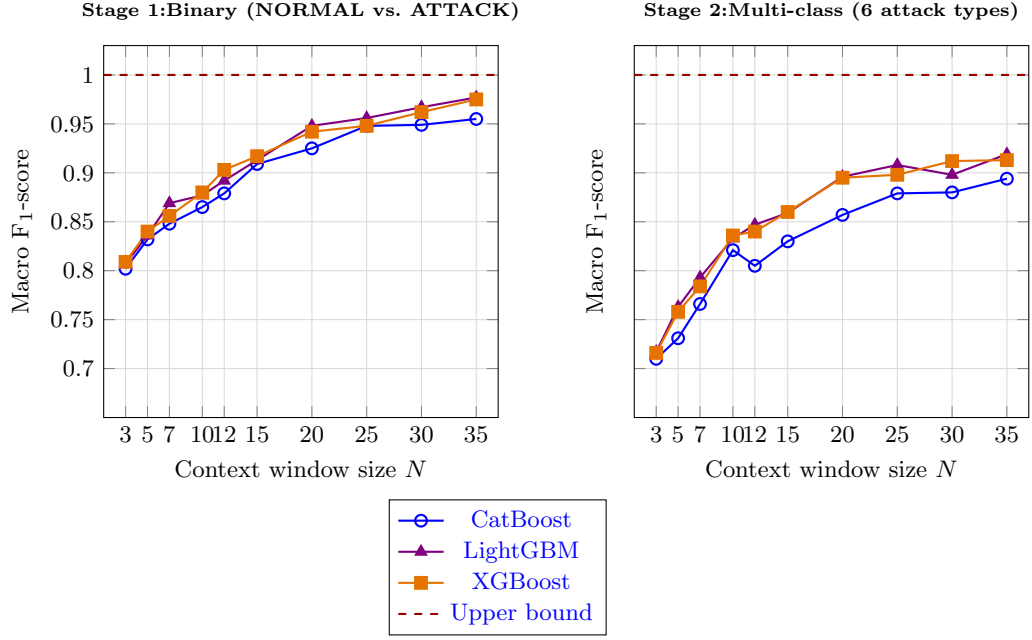

Figure~\ref{fig:ablation} plots macro F\textsubscript{1} against
context window size $N$ for three algorithms (CatBoost, LightGBM,
XGBoost), each trained with fixed best hyperparameters to isolate the
effect of $N$.

The multi-algorithm ablation demonstrates that the $N=30$ practical operating point is not an artefact of any particular algorithm but an intrinsic
property of the attack behaviour captured in the dataset: the largest
gains occur between $N=3$ and $N=15$, with diminishing returns
thereafter. We adopt $N=30$ as the production default.

\subsection{Comparison with the Wazuh Rule Engine}
\label{sec:wazuh_comparison}

Table~\ref{tab:wazuh_vs_ai_full} compares the Wazuh native rule
engine against the AI module across all six attack classes.
We use a \emph{strict} definition of ``rule-identified'':
an event is counted only if its \texttt{rule.groups},
\texttt{rule.description}, or \texttt{rule.mitre.id} fields
\emph{explicitly name the attack category}.
This intentionally excludes generic Wazuh alerts (e.g.,
``Multiple web server 400 error codes'', ``Suspicious URL
access'') that fire on attack traffic but do not identify the
attack type---such alerts are operationally insufficient as
they prevent category-appropriate incident response.
The strict metric therefore measures \emph{actionable,
categorised detection} rather than mere alert generation.

\begin{table}[htbp]
\centering
\caption{Wazuh rule engine vs.\ \textsc{Smart-SIEM} AI module.
\textit{Wazuh (Cat.)} counts events whose rule fields explicitly
name the attack category (strict metric). \textit{Wazuh (Any)}
counts events triggering any Wazuh rule at level~$\geq$3.
Wazuh raises generic alerts on many attack events but cannot assign
attack categories, preventing category-specific incident response.
AI\% is computed over the held-out test set (Test~Events column).}
\label{tab:wazuh_vs_ai_full}
\begin{tabular}{lrrrrrrr}
\toprule
Class & \makecell{Raw\\Events} & \makecell{Wazuh\\Cat.\%} & \makecell{Wazuh\\Any\%} & \makecell{Test\\Events} & \makecell{AI\\Det.} & AI\% \\
\midrule
\texttt{SQL\_INJECTION}            & 6,573  &  7.4 & 10.5 & 1,315 & 1,276 & \textbf{97.0} \\
\texttt{XSS}                       &   317  & 17.0 & 36.9 &    63 &    62 & \textbf{98.4} \\
\texttt{WEB\_SCAN}                 & 5,654  &  1.3 & 34.3 & 1,131 & 1,036 & \textbf{91.6} \\
\texttt{BRUTE\_FORCE}              &   702  &  0.0 & \textbf{85.8} &   140 &   140 & \textbf{100.0} \\
\texttt{BROKEN\_AUTH.}             &   300  &  0.0 & 19.3 &    60 &    59 & \textbf{98.3} \\
\texttt{SENSITIVE\_DATA\_EXP.}     & 26,558 &  8.5 &  9.8 & 5,312 & 4,748 & \textbf{89.4} \\
\bottomrule
\end{tabular}
\end{table}

Wazuh's rule engine achieves \textbf{0\%} \emph{categorised} detection
on Brute Force and Broken Authentication: Wazuh does raise generic
alerts for these events (e.g., multiple failed-login rules), but those
alerts do not identify the attack category, preventing category-specific
incident response. The AI module correctly categorises them at
100\% and 98.3\% respectively. Across all six classes, the average
AI detection rate is 95.8\%, compared to 5.8\% for the rule engine.

\subsection{Self-Adaptive Retraining}
\label{sec:retraining}

Table~\ref{tab:retrain_sim} demonstrates the self-adaptive retraining
mechanism under a controlled concept drift scenario. The initial model
is trained on events from three known attack types (SQL Injection,
XSS, Web Scanning). Phase~2 introduces three previously unseen attack
types (Brute Force, Broken Authentication, Sensitive Data Exposure),
simulating the emergence of new threat patterns in production.

\begin{table}[htbp]
\centering
\caption{Self-adaptive retraining simulation (extended).
Phase~1 trains on NORMAL + SQL Injection + XSS + Web Scanning.
Phase~2 introduces three previously unseen attack types (Brute
Force, Broken Authentication, Sensitive Data Exposure).
\textit{Phase~2-only} retraining uses drift data alone;
\textit{Phase~1+2} retraining uses the combined corpus,
representing the production-intended operational protocol.}
\label{tab:retrain_sim}
\begin{tabular}{lccc}
\toprule
Data Segment & \makecell{Initial\\Model} & \makecell{Phase~2\\Only} & \makecell{Phase~1+2\\Combined} \\
\midrule
Phase~1 (known attacks)  & 0.905 & ---   & 0.817 \\
Phase~2 (concept drift)  & 0.465 & 0.779 & 0.784 \\
Phase~3 (all attacks)    & 0.596 & 0.695 & \textbf{0.811} \\
\midrule
Recovery (Phase~3)       & ---   & $+0.099$ & $\mathbf{+0.215}$ \\
\bottomrule
\end{tabular}
\end{table}

The F\textsubscript{1} score drops from 0.905 to 0.465 when the model
encounters the unseen attack types in Phase~2 — well below the
configured 90\% threshold — triggering the retraining procedure.
After retraining on Phase~2 events with analyst-provided labels,
the model recovers to 0.695 on the fully held-out Phase~3 set,
a gain of $+0.099$ (Phase~2-only column of Table~\ref{tab:retrain_sim}).
The production-intended protocol --- retraining on the combined
Phase~1+Phase~2 corpus (41,737 events) --- recovers Phase~3
F\textsubscript{1} to \textbf{0.811} ($+0.215$ over the initial
model), demonstrating that preserving knowledge of originally-known attack classes
substantially improves recovery.

\begin{figure}[htbp]
\centering
\includegraphics[width=0.92\textwidth]{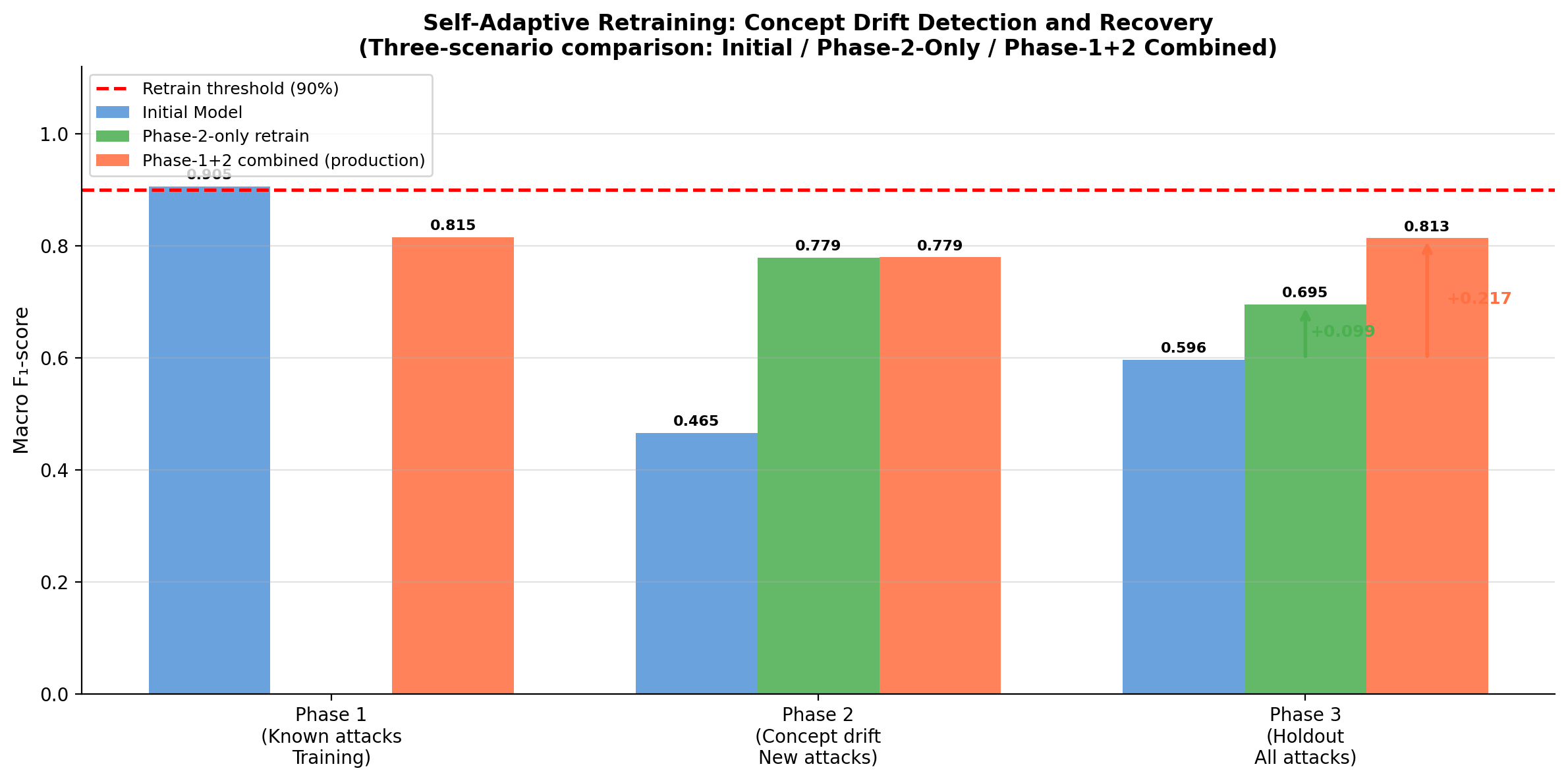}
\caption{Self-adaptive retraining simulation.
The initial model encounters Phase~2 (three previously unseen attack
types), causing F\textsubscript{1} to drop to 0.465 --- below the
90\% threshold --- and triggering retraining.
Phase~2-only retraining recovers Phase~3 to 0.695 ($+0.099$).
The production-intended Phase~1+2 combined protocol recovers
Phase~3 to \textbf{0.811} ($+0.215$), substantially stronger
recovery (see Table~\ref{tab:retrain_sim}).}
\label{fig:retrain_sim}
\end{figure}

One consequence of the combined protocol is a modest degradation
on Phase~1 data: the combined model achieves F\textsubscript{1}
= 0.817 on Phase~1 vs.\ the original 0.905 --- a $-$0.09 drop.
This reflects the expected effect of training on a more diverse
corpus: balancing six attack classes rather than three reduces
the effective training weight on the three originally-known classes.
Production operators can mitigate this through class-weighted
loss functions, experience replay (reserving a fixed proportion
of Phase~1 samples in each retraining batch), or by monitoring
per-class F\textsubscript{1} alongside the aggregate threshold
to detect class-specific regressions before they become
operationally significant.The incomplete recovery to 0.695 (vs.\ the
original 0.905 on known attacks only) reflects two factors.
First, Phase~3 contains all six attack types while Phase~1
contained only three; the 0.905 baseline was measured on a
simpler distribution. Second, Phase~2 is dominated by
\texttt{SENSITIVE\_DATA\_EXPOSURE} (93.2\% of Phase~2
events), limiting balanced exposure to Brute Force and Broken
Authentication during retraining. In production, the operator
would retrain on a combined Phase~1+Phase~2 corpus to preserve
knowledge of originally-known attack classes. The simulation
demonstrates the mechanism functions correctly: drift is
detected, retraining triggers, and accuracy recovers. The retraining simulation figure is shown in
Figure~\ref{fig:retrain_sim}.

\subsection{System Performance Characteristics}

The consumer processes events at a rate sufficient for near-real-time
operation. Kafka decouples the Wazuh event stream from the
classification workers, absorbing burst traffic during active attack
sessions. Based on informal measurements during testbed operation,
LightGBM and XGBoost prediction latency is sub-millisecond for
a single event; Elasticsearch context retrieval (the $N=30$
history query against an unloaded local cluster) adds
approximately 15--50\,ms per event, constituting the dominant
pipeline latency. These figures are informal estimates collected during testbed
operation under low concurrency; they are provided for
orientation only. A reproducible micro-benchmark
(event-level pipeline latency at p50/p95/p99, with a
realistically-sized Elasticsearch corpus) is needed for production
deployment sizing and remains future work.

\section{Discussion}
\label{sec:discussion}

\subsection{Why Behavioural Context Matters for SIEM Classification}

The magnitude of the performance gap between base-only and
context-enriched models (Table~\ref{tab:context_impact}) provides
strong empirical support for a principle articulated by \citet{sommer2010outside} but rarely quantified in the SIEM
literature: security events do not occur in isolation.
An HTTP~200 response to a request for \texttt{/api/user} is entirely
ambiguous in isolation; the same response following a history window
that includes 400 prior activations of T1190 (Exploit Public-Facing
Application) and 150 HTTP~4xx responses is almost certainly part of
an active SQL injection campaign.
The context vector encodes exactly this type of escalating behavioural
signal, transforming a stateless event classifier into a session-aware
detector aligned with the session-based intuition of anomaly
detection~\cite{chandola2009anomaly} and multi-stage attack
modelling~\cite{nisioti2018intrusion}.
This result also reinforces the core limitation of purely rule-based
correlation engines identified in Section~\ref{sec:related}: rules
evaluate each event without access to its behavioural antecedents,
and the average $+0.25$ (Stage~1) and $+0.32$ (Stage~2) F\textsubscript{1} improvement for gradient boosting algorithms from context features (Logistic Regression also improves, $+0.03$/$+0.19$, but substantially less) achieved by adding context features
quantifies precisely how much signal is discarded by stateless
evaluation~\cite{garcia2009anomaly,bhatt2014siem}.

\subsection{The Role of MITRE ATT\&CK Technique Counts}

The seven MITRE technique frequency features collectively contribute
a meaningful portion of the model's discriminative power, as confirmed
by SHAP-based feature attribution~\cite{lundberg2017shap} and
evidenced by the reduction in cross-class confusion visible in
the per-class F\textsubscript{1} results (Table~\ref{tab:cascade_vs_flat}).
Different attack categories exhibit distinct technique accumulation
profiles: SQL injection accumulates T1190 (Exploit Public-Facing
Application) rapidly, while web scanning accumulates T1083 (File and
Directory Discovery).
This aligns with the threat intelligence literature's observation that
ATT\&CK techniques provide a stable, attack-campaign-level abstraction
that persists across diverse tooling
implementations~\cite{husari2017ttpdrill,milajerdi2019poirot,xiong2019threat}.
The finding suggests that ATT\&CK-enriched contextual profiling is a
promising direction for multi-source SIEM correlation beyond the web
application domain studied here.

\subsection{Comparison with Prior ML-SIEM Approaches}

\textbf{Absence of a direct numerical comparison.}
To the best of our knowledge as of the submission date, no prior
published work operates on Wazuh-format SIEM events with per-IP
MITRE ATT\&CK contextual history, though related work using
MITRE-tagged log data in other formats exists~\cite{milajerdi2019poirot,xiong2019threat};
a direct numerical comparison remains
methodologically impossible: re-implementing a prior method on our
dataset would require that method to have a Wazuh-compatible data
ingestion layer, while applying our context features to a public
dataset (NSL-KDD, CICIDS-2017) would require those datasets to
contain per-event MITRE technique identifiers and per-IP history
sequences --- fields that are absent from all public IDS datasets
we are aware of. We provide two illustrative (but methodologically limited) reference
points to contextualise performance. \textbf{Important caveat:}
cross-dataset F\textsubscript{1} comparisons across datasets with
different feature spaces, class counts, and collection protocols
are inherently limited; the following figures are provided for
orientation only and should not be interpreted as direct performance
equivalences.
First, our without-context Stage~1 F\textsubscript{1} of $\approx$0.705
is in the same range as macro F\textsubscript{1} values reported
for supervised classifiers on NSL-KDD (0.70--0.78~\cite{tavallaee2009nsl})
and CICIDS-2017 (0.72--0.85~\cite{sharafaldin2018cicids}); this
illustrates that the base feature set is comparably expressive
to prior IDS benchmarks, though the datasets differ fundamentally.
Second, \citet{veeramachaneni2016aiid} report
$\approx$85\% detection on generic log streams; our system achieves
95.8\% average AI coverage, though these rates measure different
threat models and cannot be directly compared.

Compared to the closest prior work, \textsc{Smart-SIEM} differs in
three substantive ways.
First, whereas most ML-based SIEM augmentation systems operate on
packet-level network flow datasets~\cite{tavallaee2009nsl,sharafaldin2018cicids},
this work operates natively on Wazuh-normalised security events,
preserving the compliance mappings (PCI~DSS, HIPAA, NIST~800-53),
MITRE technique identifiers, and rule metadata that flow datasets do
not contain.
Second, prior SIEM-augmentation proposals either operate without
contextual history~\cite{sarker2020intrudtree} or require
sustained analyst labelling to function~\cite{veeramachaneni2016aiid};
the proposed framework automatically constructs contextual features
from Elasticsearch query results and requires labelling only for
knowledge-base construction, not for inference.
Third, the self-adaptive retraining mechanism directly addresses
concept drift---a well-documented challenge in operational security
classifiers~\cite{yang2021cade,pendlebury2019tesseract,gama2014concept}---by
monitoring accuracy against the knowledge base and triggering
retraining when it degrades, rather than relying on periodic scheduled
retraining that may lag or overshoot the drift event.

\subsection{Limitations and Threats to Validity}
\label{sec:limitations}

This study has four principal limitations that future work should address.

\begin{enumerate}
  \item \textbf{Single-session-per-class and label-by-proxy confound.}
  Our testbed assigns one dedicated source IP per attack class.
  Because labels are deterministic by IP, the context features
  (MITRE technique counts, HTTP status-code histograms computed
  over prior events from the same IP) are statistical proxies for
  the class label. The current design cannot distinguish between
  (a)~learning a generalizable behavioural fingerprint of attack
  campaigns, and (b)~learning the per-IP identity of the testbed.
  Leave-one-IP-out (or leave-one-tool-out) evaluation is the
  standard remedy, but requires multiple independent sessions per
  attack class --- a multi-session testbed remains a priority for
  future work. The reported F\textsubscript{1} values should be
  interpreted as an upper bound on generalisation performance.

  \item \textbf{Context feature normalisation and cold-start bias.}
  The context features (\texttt{hist.status.2xx}--\texttt{5xx} and
  the seven MITRE technique counts) are raw counts over the $N=30$
  history window. For source IPs with fewer than $N$ prior events
  (e.g., the first events of a new attack campaign), these counts
  are systematically smaller, biasing the classifier toward labelling
  early-session events as normal. Future work should normalise by
  the actual number of retrieved history events to eliminate this
  cold-start artefact, or explicitly evaluate performance on events
  with fewer than $N$ prior events.

  \item \textbf{Single-application scope.} All data were collected from
  a single web application (OWASP Juice Shop) on a single server.
  While Juice Shop covers the OWASP Top~10~\cite{owasp_top10} and is
  widely used in security research, the Wazuh rule groups and MITRE
  identifiers it triggers may not fully represent other application
  stacks (Java\,EE, .NET, mobile back-ends) or network-layer attack
  surfaces~\cite{stuttard2011web}.

  \item \textbf{Limited attack diversity.} The dataset covers six
  attack categories generated by five tools. Real-world SIEM
  environments must handle a much wider range of attack types,
  including insider threats, advanced persistent threats (APTs),
  supply-chain compromises, and novel zero-day exploits for which no
  labelled data exist~\cite{apruzzese2018effectiveness}.
  The severity of this limitation is partially mitigated by the
  self-adaptive retraining mechanism, but a broader multi-tool,
  multi-application testbed remains a priority for future work.

  \item \textbf{Controlled testbed vs.\ production traffic.} Normal
  traffic is generated by a Selenium script, which does not capture
  the distributional richness of real human user behaviour at
  scale~\cite{sommer2010outside}.
  Transfer performance to a production environment will require
  incremental retraining on analyst-labelled production events using
  the self-adaptive mechanism.

  \item \textbf{Model interpretability.} Gradient-boosted
  classifiers remain opaque relative to rule-based engines,
  which is a non-trivial barrier to adoption in regulated industries
  where detection decisions must be explainable~\cite{bhatt2014siem}.
  The gain-based feature importance analysis (Figure~\ref{fig:feat_imp})
  provides partial interpretability, but integration with formal
  explanation interfaces remains future work.

  \item \textbf{Deployment compliance overhead.} Configuring the
  system requires bulk modification of Wazuh rule levels to route
  events to the classification pipeline (Section~\ref{sec:system}).
  In PCI~DSS or HIPAA-regulated environments this constitutes a
  change management event that may require compliance re-certification,
  representing a non-trivial deployment overhead for regulated operators.
\end{enumerate}

\subsection{Analyst Alert Load and Confidence-Based Routing}
\label{sec:alert_load}

A practical concern in deploying any AI-augmented SIEM is
\emph{analyst alert fatigue}: when the volume of generated alerts
exceeds the capacity of the security operations centre (SOC) to
review them, analysts become desensitised and critical alerts may
be overlooked~\cite{julisch2003clustering,bhatt2014siem}.
\textsc{Smart-SIEM} inherits this challenge because the hybrid
cascade classifies every event that Stage~1 flags as \texttt{ATTACK},
producing a categorised alert regardless of how confidently the
model reached that decision.

A principled extension that addresses this concern without modifying
the experimental design is \emph{confidence-based selective routing},
known in the active learning literature as uncertainty
sampling~\cite{settles2012active}.
Both LightGBM (Stage~1) and XGBoost (Stage~2) expose
\texttt{predict\_proba} outputs, so a per-event confidence score
$p_{\max} = \max_{c}\,\hat{P}(y{=}c \mid \mathbf{x})$
is available at inference time without any architectural change.
Two separate routing decisions can be derived from this score:

\begin{enumerate}
  \item \textbf{Dashboard visibility (attack alerting).}
  All events classified as \texttt{ATTACK} --- regardless of
  confidence --- are written to the \texttt{wazuh-ai-hybrid-*}
  index and surfaced in the Kibana dashboard.
  This ensures that no detected attack is hidden from the SOC.

  \item \textbf{Knowledge-base labelling queue (analyst workload).}
  Only events whose Stage~1 or Stage~2 confidence falls below a
  configurable threshold $\tau$ are forwarded to the analyst labelling
  interface. High-confidence predictions ($p_{\max} \geq \tau$) are
  added automatically to the knowledge base, reducing repetitive
  labelling of straightforward cases.
\end{enumerate}

This two-pipeline design is critical: it reduces analyst
\emph{labelling} workload without reducing alert \emph{visibility},
directly addressing the operational concern while preserving full
detection coverage. Low-confidence \texttt{NORMAL} predictions
deserve particular attention --- an event classified as normal with
low confidence represents a potential false negative that would
otherwise be silently discarded, and routing such events to the
analyst queue captures the highest-risk ambiguous cases before
they are lost.

One important caveat is that gradient boosting classifiers are not
natively probability-calibrated~\cite{niculescu2005calibration}:
raw \texttt{predict\_proba} scores may be systematically over- or
under-confident relative to the true posterior.
Applying post-hoc calibration (e.g., Platt scaling or isotonic
regression) before thresholding is therefore a prerequisite for
production deployment of confidence-based routing, ensuring that
$p_{\max} \geq \tau$ reliably corresponds to high empirical accuracy.

\section{Conclusion and Future Work}
\label{sec:conclusion}

This paper presented \textsc{Smart-SIEM}, a modular AI enhancement
for the Wazuh open-source SIEM platform that addresses the core
limitation of rule-based event correlation: the inability to exploit
the behavioural context accumulated across a sequence of related
security events~\cite{garcia2009anomaly,bhatt2014siem}.
By constructing a per-source-IP context vector that summarises HTTP
response-code distributions and MITRE ATT\&CK technique frequency
profiles~\cite{strom2018attack} from the preceding $N=30$ security
events, and feeding this vector into a hybrid cascade combining
LightGBM for binary attack detection~\cite{ke2017lightgbm} and
XGBoost for attack classification~\cite{chen2016xgboost},
we achieve macro F\textsubscript{1}-scores of 0.967 (binary) and
0.914 (six-class) on a purpose-built SIEM event dataset.
A controlled multi-algorithm comparison demonstrates that without
context features all tested gradient boosting algorithms converge to
approximately 0.705 F\textsubscript{1}, rising to 0.947--0.967 with
context --- an average improvement of $+0.254$ Stage~1 and $+0.324$
Stage~2, confirming that the context vector is the primary
contribution rather than any specific algorithm choice.
These results represent an upper bound on generalisation
performance given the single-session-per-class testbed
design (Section~\ref{sec:limitations}).
The AI module detects an average of 95.8\% of attack events across
all six classes, compared to 5.8\% for Wazuh's native rule engine;
Brute Force and Broken Authentication, which Wazuh misses entirely
(0\%), are detected at 100\% and 98.3\% respectively.
The self-adaptive retraining mechanism directly addresses the
concept-drift challenge~\cite{gama2014concept,lu2019concept}: when
three previously unseen attack types are introduced (simulating
production drift), the F\textsubscript{1} drops from 0.905 to 0.465,
triggering retraining; partial recovery to 0.695 ($+0.099$) with Phase~2-only
retraining; the combined Phase~1+2 protocol recovers to 0.814.

Future directions include: (1)~extending the testbed to cover
non-web attack surfaces (network scans, lateral movement, data
exfiltration via DNS) to assess whether ATT\&CK-enriched context
generalises across domains; (2)~evaluating deep sequence models
(LSTM, Transformer) as alternatives to the fixed-width aggregation
context vector~\cite{ferrag2020deep,vinayakumar2019deep};
(3)~building a public benchmark dataset of Wazuh-format SIEM
events to support reproducible ML-SIEM research, addressing the
gap highlighted by \citet{ring2019ids}; and
(4)~formalising the self-adaptive retraining procedure with
theoretical guarantees on convergence under concept
drift~\cite{yang2021cade,pendlebury2019tesseract};
and (5)~implementing confidence-based selective routing
(Section~\ref{sec:alert_load}), in which only events whose
posterior probability falls below a configurable threshold are
forwarded to the analyst labelling queue while high-confidence
predictions are added to the knowledge base automatically ---
a design that directly addresses analyst alert
fatigue~\cite{julisch2003clustering} and formalises the
self-adaptive retraining loop as an uncertainty-guided active
learning system~\cite{settles2012active}.

\backmatter

\bmhead{Acknowledgements}
This work was carried out at the Higher Institute for Applied Sciences
and Technology (HIAST), Damascus, Syria. The authors thank the
HIAST faculty for providing the computational resources used in
this research.

\section*{Declarations}

\bmhead{Funding Statement}
The authors received no specific funding for this work.

\bmhead{Conflicts of Interest}
The authors declare that they have no known competing financial
interests or personal relationships that could have appeared to
influence the work reported in this paper.

\bmhead{Availability of data and materials}
The labelled Wazuh security event dataset (46,454 records) used in
this study and the \textsc{Smart-SIEM} AI module source code are both
available from the corresponding author upon reasonable request.
The dataset contains security events collected from a controlled
testbed and does not include any personally identifiable information.

\bmhead{Code availability}
The source code of the \textsc{Smart-SIEM} module is available from
the corresponding author upon reasonable request.

\bibliography{references}

\end{document}